\documentclass[prb,twocolumn,superscriptaddress,showpacs]{revtex4-1}

\pdfoutput=1

\usepackage{graphicx}
\usepackage{amsmath,amssymb}

\usepackage{color}
\usepackage{amssymb}
\usepackage{braket}

\usepackage{hyperref}
\hypersetup{colorlinks=true,breaklinks,linkcolor=blue,urlcolor=blue,citecolor=blue}

\def\mathi{\mathrm i}

\newcommand{\bS}{\ensuremath{\boldsymbol{S}}}
\newcommand{\bU}{\ensuremath{\boldsymbol{U}}}
\newcommand{\bV}{\ensuremath{\boldsymbol{V}}}

\newcommand{\bK}{\ensuremath{\boldsymbol{K}}}

\newcommand{\wmax}{\ensuremath{{\omega_\mathrm{max}}}}

\newcommand{\kF}{\ensuremath{k^\mathrm{F}}}
\newcommand{\kB}{\ensuremath{k^\mathrm{B}}}

\newcommand{\Nag}{\ensuremath{N^{\alpha,\gamma}}}
\newcommand{\NFIR}{\ensuremath{N^\mathrm{F,IR}}}
\newcommand{\NBIR}{\ensuremath{N^\mathrm{B,IR}}}
\newcommand{\NaIR}{\ensuremath{N^{\alpha,\mathrm{IR}}}}
\newcommand{\NFL}{\ensuremath{N^\mathrm{F,L}}}

\newcommand{\NaL}{\ensuremath{N^{\alpha,\mathrm{L}}}}

\newcommand{\NFs}{\ensuremath{M^\mathrm{F}}}
\newcommand{\NBs}{\ensuremath{M^\mathrm{B}}}
\newcommand{\Nas}{\ensuremath{M^{\alpha}}}

\newcommand{\slB}{\ensuremath{s_l^\mathrm{B}}}
\newcommand{\sla}{\ensuremath{s_l^\alpha}}

\begin{document}
\title{Performance analysis of a physically constructed orthogonal representation of imaginary-time Green's function}
\author{Naoya Chikano}
\affiliation{Department of Physics, Saitama University, 338-8570, Japan}

\author{Junya Otsuki}
\affiliation{Department of Physics, Tohoku University, Sendai 980-8578, Japan}

\author{Hiroshi Shinaoka}
\email{shinaoka@mail.saitama-u.ac.jp}
\affiliation{Department of Physics, Saitama University, 338-8570, Japan}

\date{\today}

\begin{abstract}
The imaginary-time Green's function is a building block of various numerical methods for correlated electron systems.
Recently, it was shown that a model-independent compact orthogonal representation of the Green's function can be constructed by decomposing its spectral representation.
We investigate the performance of this so-called \textit{intermediate representation} (IR) from several points of view.
First, we develop an efficient algorithm for computing the IR basis functions of arbitrary high degree.
Second, for two simple models,
we study  the number of coefficients required to represent the Green's function within a given tolerance.
We show that the number of coefficients grows only as $O(\log \beta)$ for fermions,
and converges to a constant for bosons as temperature $T=1/\beta$ decreases.
Third, we show that this remarkable feature is ascribed to the properties of the physically constructed basis functions.
The fermionic basis functions on the real-frequency axis have features whose width is scaled as $O(T)$, which are consistent with the low-$T$ properties of quasiparticles in a Fermi liquid state.
On the other hand, the properties of the bosonic basis functions are consistent with those of spin/orbital susceptibilities at low $T$.
These results demonstrate the potential wide applications of the IR to calculations of correlated systems.
\end{abstract}

\maketitle
\section{Introduction}
Many theoretical frameworks for describing correlated electrons are based on imaginary-time Green's function theories.
Examples include the dynamical mean-field theory (DMFT)~\cite{Georges:1996un}, various diagrammatic methods~\cite{AGD,Aryasetiawan:1998wz,Metzner:2012jv},
and quantum Monte Carlo (QMC) methods~\cite{Gull:2011jda,Sandvik:1991ht,Rombouts:1999tz}.
Recently, much effort has been made to study correlated electron materials by means of these methods in combination with the density functional theory.\cite{Biermann02,Kotliar:2006fl,Platt:2011iz}
Such calculations for realistic systems sometimes suffer from large computational cost and massive memory consumptions.
This problem becomes particularly severe at low temperature as the data size of the Green's function increases.
It is thus practically important to develop a compact representation of the Green's function.

In the context of DMFT,
it was recently proposed to represent the imaginary-time dependence of the single-particle Green's functions, $G(\tau)$, in terms of Legendre polynomials.\cite{Boehnke:2011dd}
This yields exponentially decaying expansion coefficients, 
while the ordinary Matsubara representation $G(i\omega_n)$ exhibits a power-law decay.
This technique has been applied to efficient quantum Monte Carlo (QMC) measurements~\cite{Huang:2015cc,Seth:2015uq,Shinaoka:2017cpa}, solution of Bethe-Salpeter equations~\cite{Boehnke:2011dd}, and some quantum chemistry calculations.\cite{Kananenka:2016cfa,Rusakov:2016eu}
The Legendre representation however still suffers from slow convergence of expansion coefficients in low-temperature regimes of physical interest.

Some of the authors and co-workers have recently proposed a physically motivated compact representation for $G(\tau)$.\cite{Shinaoka:2017ix}
The formalism is based on the fact that extracting the spectral function $\rho(\omega)$ from $G(\tau)$ is an ill-posed problem (analytical continuation).
This means that $G(\tau)$ has less information than $\rho(\omega)$ and hence is compressible without loss of relevant information.
They demonstrated that high compression of $G(\tau)$ is indeed achieved by a basis derived as the ``intermediate representation (IR)" between imaginary-time and real-frequency domains.
An interesting finding was that the IR basis functions converge to Legendre polynomials in the high-temperature limit of a control parameter $\Lambda$ (definition is given later).
Away from this limit, the IR functions constitute a non-polynomial orthogonal basis set which yields a faster convergence of expansion coefficients than the Legendre representation.
A succeeding study further showed that the IR basis functions can be used to construct a compact representation of the two-particle Green's functions.\cite{shinaoka-unpublished}

In this paper, we present a systematic study on the performance of the compact representation using the IR basis.
In particular, we investigate how the compactness of the IR depends on temperature $T$ in comparison with the conventional Legendre representation. 
We show that the number of required basis functions increases only logarithmically against $1/T$ for fermions, while it saturates to a constant for bosons
at low temperature.
This scaling clearly indicates a qualitative superiority of the IR basis over the Legendre representation which yields a power-law increase of the basis functions in the expansion.
By revealing the features of the IR basis functions,
we provide an insight into how the IR provides a compact representation of the Green's functions.

The remainder of this paper is organized as follows.
In the next section, we first review the Legendre representation and the IR to establish notations.
We also outline an efficient method for computing the IR basis functions.
Section III presents results of performance analysis of these representations using simple models.
The general properties of the IR basis functions are analyzed in Sec. VI.
Section V presents a summary and conclusions.

\section{Orthogonal representations}\label{sec:basis}
We review the IR for the single-particle Green's function.
We first give a brief description on the Legendre representation,
and then introduce the IR.

\subsection{Legendre polynomial representation}
We consider the single-particle imaginary-time Green's function $G^\alpha(\tau)$.
The superscript $\alpha$ specifies statistics: $\alpha=\mathrm{F}$ for fermion and $\alpha=\mathrm{B}$ for boson.
The Matsubara Green's function is given by
\begin{align}
	G^{\alpha}(i\omega_n) &= \int_0^\beta d\tau e^{i\omega_n \tau} G^\alpha(\tau),
\end{align}
where $\omega_n = (2n+1)\pi/\beta$ for fermion and $\omega_n = 2n\pi/\beta$ for boson.
$G^\alpha(i\omega_n)$ has a power-law tail at high frequencies,
which prevents compact representation of $G^\alpha(i\omega_n)$ in practical applications.

Boehnke \textit{et al.} proposed to expand $G^\alpha(\tau)$ in terms of Legendre polynomials~\cite{Boehnke:2011dd}.
Since $G^\alpha(\tau)$ is continuous and smooth in the interval of $[0,\beta]$, one can represent $G^\alpha(\tau)$ as
\begin{align}
	G^\alpha(\tau) &= \sum_{l\le 0}\frac{\sqrt{2l+1}}{\beta} P_l(x(\tau))G_l^{\alpha,\mathrm{L}},\label{eq:LG-exp}\\
	G_l^{\alpha,\mathrm{L}} &= \sqrt{2l+1} \int_0^\beta d \tau P_l(x(\tau)) G(\tau),\label{eq:LG-exp-coeff}
\end{align}
where $x(\tau)\equiv 2\tau/\beta -1$ and $P_l(x)$ is the Legendre polynomial of degree $l$.
It was demonstrated that the expansion coefficients $G_l^{\alpha,\mathrm{L}}$ decays exponentially.
In continuous-time QMC calculations, $G_l^{\alpha,\mathrm{L}}$ can be measured directly, which avoids truncation errors in the Matsubara-frequency representation. 
However, expansions in terms of Legendre polynomials become inefficient as $T$ decreases, since $G^\alpha(\tau)$ varies abruptly near $\tau=0$ and $\beta$ at low temperatures.

\subsection{Intermediate representation of imaginary-time and real-frequency domains}
\subsubsection{Definition of basis functions}
The IR basis is derived from the spectral (Lehmann) representation, which connects $G^{\alpha}(\tau)$ with the spectral function $\rho^\alpha(\omega)$ on the real-frequency axis
\begin{align}
	G^{\alpha}(\tau) &= \int_{-\infty}^\infty d \omega~ \rho^\alpha(\omega) K^\alpha(\tau, \omega), \label{eq:Gtau}
\end{align}
where we take $\hbar = 1$.
Here, we define the spectral function as
\begin{align}
	\rho^\alpha(\omega) &= -\frac{1}{\pi\omega^{\delta_{\alpha,\mathrm{B}} }} \mathrm{Im} G^\alpha(\omega + \mathi 0).\label{eq:rho}
\end{align}
The kernel $K^{\alpha}(\tau,\omega)$ reads
\begin{align}
	K^\alpha(\tau, \omega) &\equiv \omega^{\delta_{\alpha, \mathrm{B}}} \frac{e^{-\tau\omega}}{1 \pm e^{-\beta \omega}}.\label{eq:K}
\end{align}
The extra $\omega$'s for boson in Eqs.~(\ref{eq:rho}) and (\ref{eq:K}) was introduced to avoid a singularity of the kernel at $\omega=0$.
Note that the physical spectrum for boson is given by $\omega \rho^\mathrm{B}(\omega)$.
The difficulty in analytical continuation arises from the \textit{infamous} nature of the kernel: It filters out most of the information contained in $\rho^\alpha(\omega)$ (e.g., features at high $\omega$).
Namely, $G^\alpha(\tau)$ contains less information than $\rho^\alpha(\omega)$~\cite{Otsuki:2017er},
which implies the existence of a highly compact representation of $G^\alpha(\tau)$ \textit{without} loss of relevant information.

The IR basis functions, $\{U^\alpha_l(\tau)\}$ and $\{V^\alpha_l(\omega)\}$, are defined by the decomposition
\begin{align}
	K^{\alpha}(\tau, \omega) &= \sum_{l=0}^{\infty} S^{\alpha}_l U^{\alpha}_l(\tau) V^{\alpha}_l(\omega),\label{eq:kernel-exp}
\end{align}
in the intervals of $[0,\beta]$ and $[-\wmax,\wmax]$.
Here, we introduced a cutoff frequency $\wmax$.
These two basis sets are orthonormalized in these intervals, respectively.
This decomposition can be regarded as the continuous limit of the singular value decomposition (SVD) of a matrix representation of $K^{\alpha}(\tau, \omega)$.
The coefficients $S_l^\alpha$ in Eq.~(\ref{eq:kernel-exp}) correspond to singular values in the SVD, which vanish exponentially as $l\rightarrow +\infty$.
Note that $U_l^\alpha(\tau)$ and $V_l^\alpha(\omega)$ are related as
\begin{align}
	S^\alpha_l U^\alpha_l(\tau) &= \int_{-\wmax}^\wmax d\omega K^\alpha(\tau, \omega) V^\alpha_l(\omega).\label{eq:integral-UV}
\end{align}

\subsubsection{Expansion of single-particle Green's function}
The central idea of the IR is to expand $G^\alpha(\tau)$ in terms of  the complete basis set $\{U^{\alpha}_l(\tau)\}$ as
\begin{align}
	G^{\alpha}(\tau) &= \sum_{l=0}^\infty  G_l^{\alpha, \mathrm{IR}} U_l^{\alpha}(\tau)\label{eq:IR-decomp}.
\end{align}
A fast decay of the expansion coefficients $G_l^{\alpha, \mathrm{IR}}$ is shown as follows.
We first assume that the spectral function $\rho^\alpha(\omega)$ is bounded in $[-\Omega, \Omega]$.
When the domain of the $\{V^\alpha_l(\omega)\}$ covers all region of $\rho^\alpha(\omega)$, namely, $\wmax \ge \Omega$,
a substitution of Eq.~(\ref{eq:kernel-exp}) into Eq.~(\ref{eq:Gtau}) leads to
\begin{align}
	G^{\alpha, \mathrm{IR}}_l = - S^{\alpha}_l \rho^{\alpha}_l,\label{eq:gl}
\end{align}
where $\rho^{\alpha}_l$ denotes the expansion coefficients of $\rho^\alpha(\omega)$ defined by
\begin{align}
	\rho_l^{\alpha} \equiv \int_{-\wmax}^{\wmax} d \omega \rho^\alpha(\omega) V^\alpha_l(\omega).\label{eq:rhol}
\end{align}
Equation~(\ref{eq:gl}) clearly shows that $G^{\alpha, \mathrm{IR}}_l$ decays at least as fast as $s_l^{\alpha}$.

Equation (\ref{eq:gl}) also indicates that 
small statistical errors in $G^{\alpha, \mathrm{IR}}_l$ are amplified when reconstructing the spectral function $\rho^\alpha(\omega)$ from QMC data. We refer the interested reader to Ref.~\onlinecite{Otsuki:2017er} for more detailed discussion on analytical continuation.

\subsubsection{Dimensionless parameter for basis functions}
The singular values $S_l^\alpha$ and the shapes of basis functions depend on $\wmax$ and $\beta$ through the dimensionless variable $\Lambda \equiv \beta\wmax$.\cite{Shinaoka:2017ix}
To see this, we change variables $\tau$ and $\omega$ into $x \equiv 2\tau/\beta -1$ and $y \equiv \omega/\wmax$.
Then, the kernels read
\begin{align}
\kF(x, y)&\equiv \frac{e^{-\frac{\Lambda}{2} x y}}{2\cosh(\frac{\Lambda}{2} y)},\label{eq:kernel-F-xy}\\
\kB(x, y)&\equiv y \frac{e^{-\frac{\Lambda}{2} x y}}{2\sinh(\frac{\Lambda}{2} y)},\label{eq:kernel-B-xy}
\end{align}
up to a constant.
Similarly to $K^\alpha(\tau,\omega)$,
one can decompose $k^\alpha(x,y)$ as 
\begin{align}
	k^\alpha(x, y) &= \sum_{l=0}^{\infty} s^\alpha_l u^\alpha_l(x) v^\alpha_l(y),\label{eq:kernel-xy-exp}
\end{align}
under the orthonormal conditions in the interval $[-1,1]$.
The singular values $s^\alpha_l$ in Eq.~(\ref{eq:kernel-xy-exp}) are identical to those in Eq.~(\ref{eq:kernel-exp}) up to a constant factor which is independent of $l$.
One can also show the relations
\begin{align}
    U_l^\alpha(\tau) &= \sqrt{\frac{2}{\beta}} u_l^\alpha(2\tau/\beta-1),\label{eq:Ul}\\
    V_l^\alpha(\omega) &= \sqrt{\frac{1}{\wmax}} v_l^\alpha(\omega/\wmax).\label{eq:Vl}
\end{align}

The previous study showed that $u_l^\alpha(x)$ and $v_l^\alpha(y)$ converge to $P_l(x)$ up to a normalization factor as $\Lambda \rightarrow 0$~\cite{Shinaoka:2017ix}.
For $\Lambda > 0$, each of them constitutes a non-polynomial orthogonal basis set.

\subsubsection{Technical details}
In practical calculations, the decompoision in Eq.~(\ref{eq:kernel-xy-exp}) can be performed by solving the integral equation
\begin{align}
	s^\alpha_l u^\alpha_l(x) &= \int_{-1}^1 dy k^\alpha(x, y) v^\alpha_l(y).\label{eq:int-solve}
\end{align}
In the present study, we solved this equation numerically in the continuous limit.
This was achieved by expanding $u^\alpha_l(x)$ and $v^\alpha_l(y)$ in terms of piecewise polynomials.
To compute basis functions for small singular values with controlled accuracy,
we used arbitrary precision arithmetic.
We refer the interested reader to Appendix~\ref{appendix:method} for more details.

\section{Analysis based on model Green's function}\label{sec:model-analysis}
\subsection{Model and method}
\begin{figure}
	\begin{flushleft}
	\hspace{4em}	(a) Gapless model
	\end{flushleft}
	\centering
	\hspace{0.5cm}\includegraphics[width=0.275\textwidth,clip]{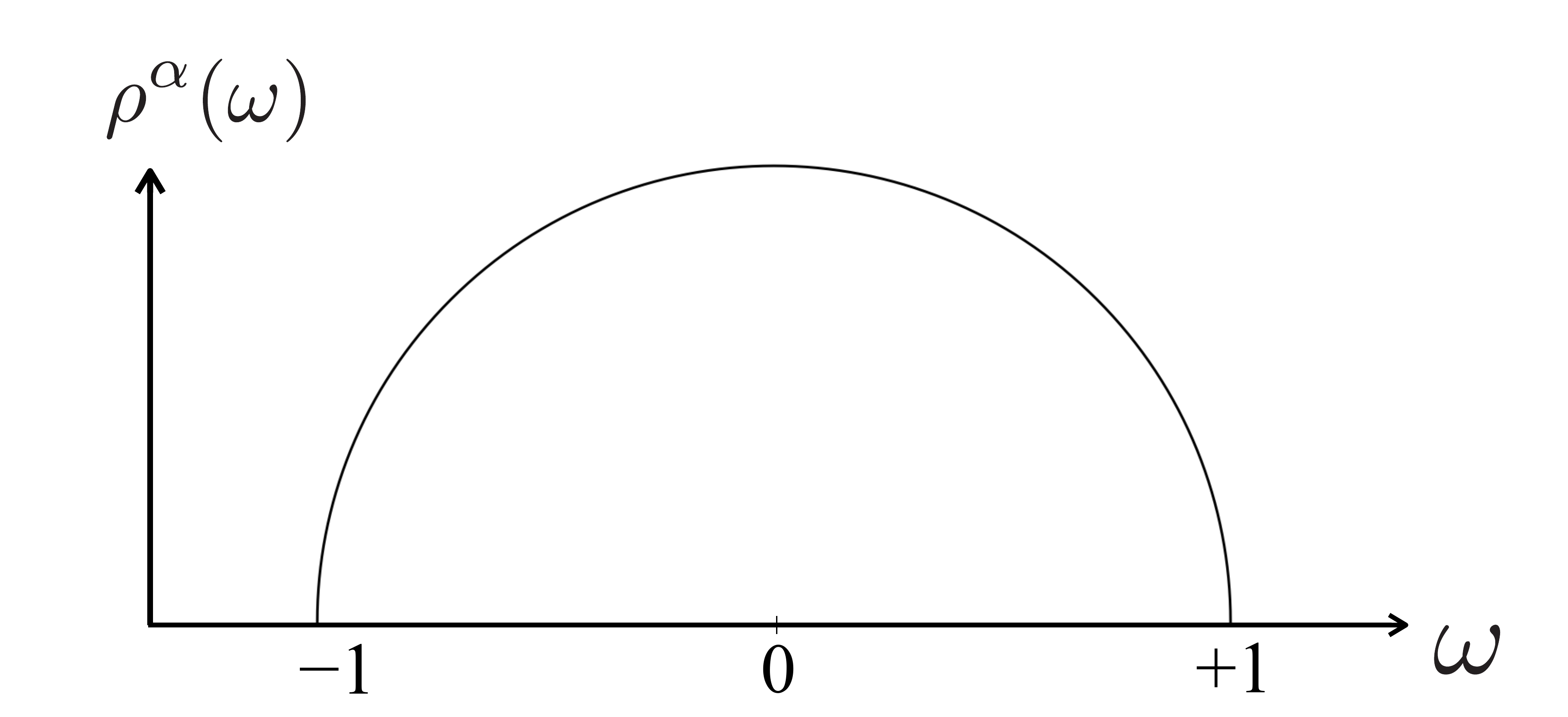}
	\includegraphics[width=0.35\textwidth,clip]{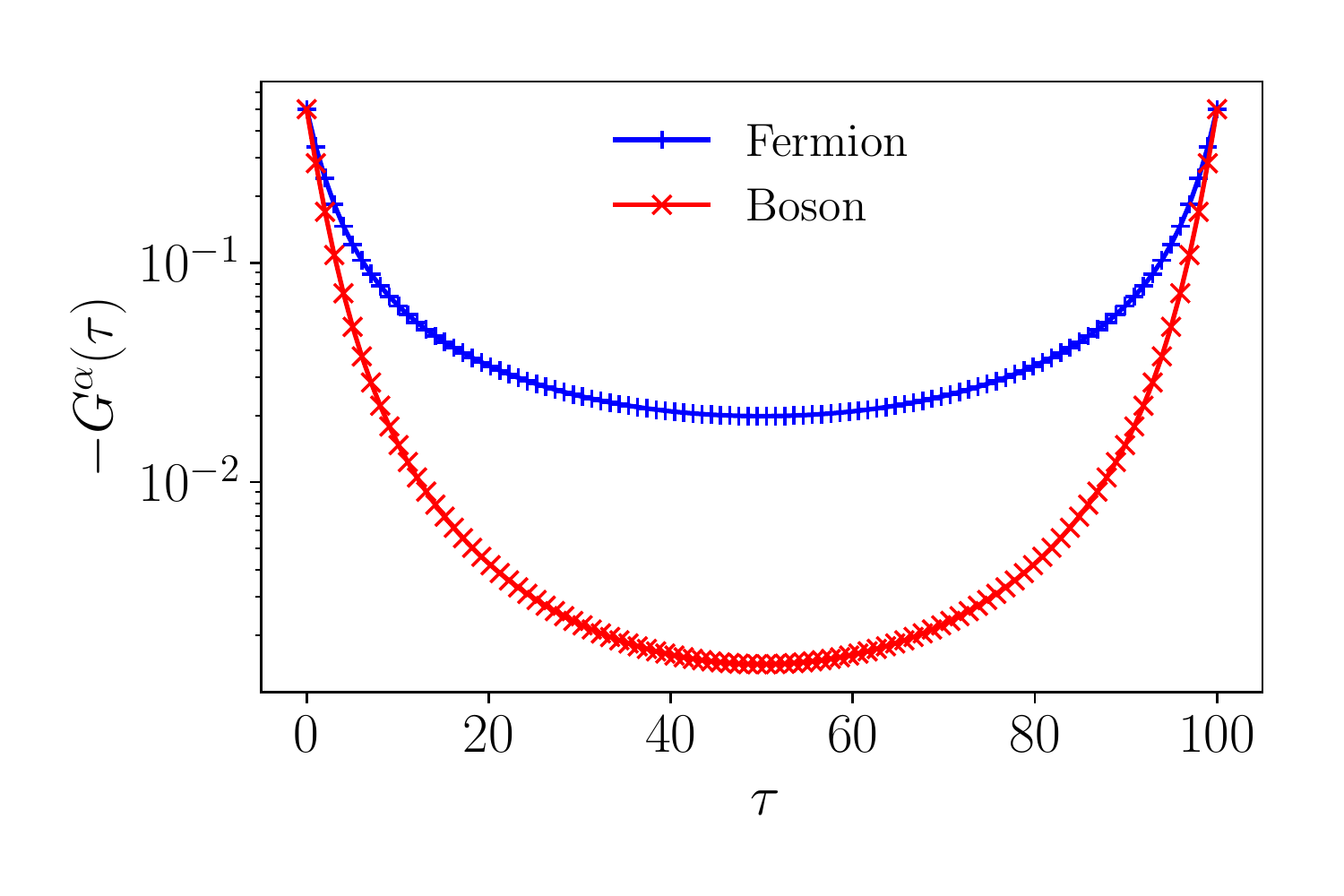}
	\begin{flushleft}
	\hspace{4em}	(b) Insulating model
	\end{flushleft}
	\centering
	\hspace{0.5cm}\includegraphics[width=0.275\textwidth,clip]{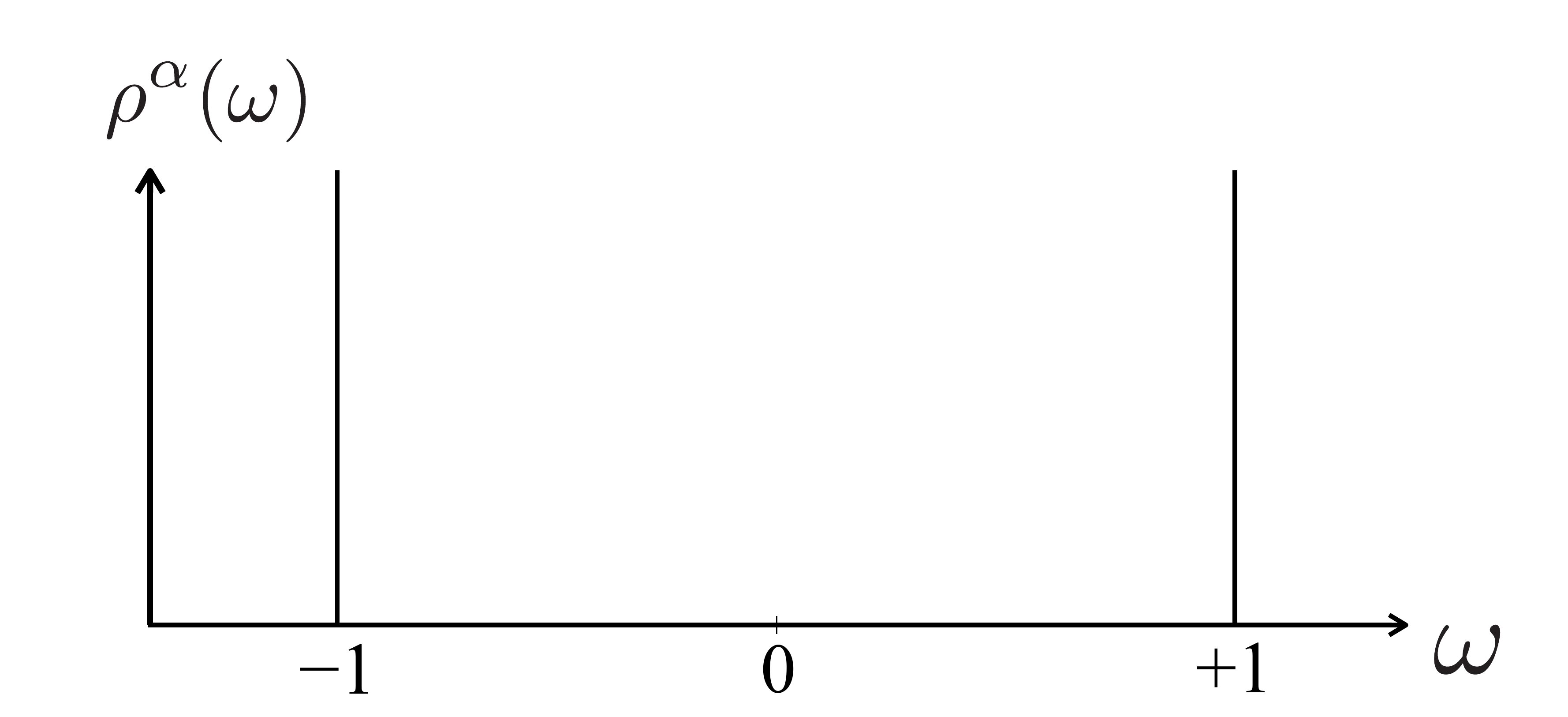}
	\includegraphics[width=0.35\textwidth,clip]{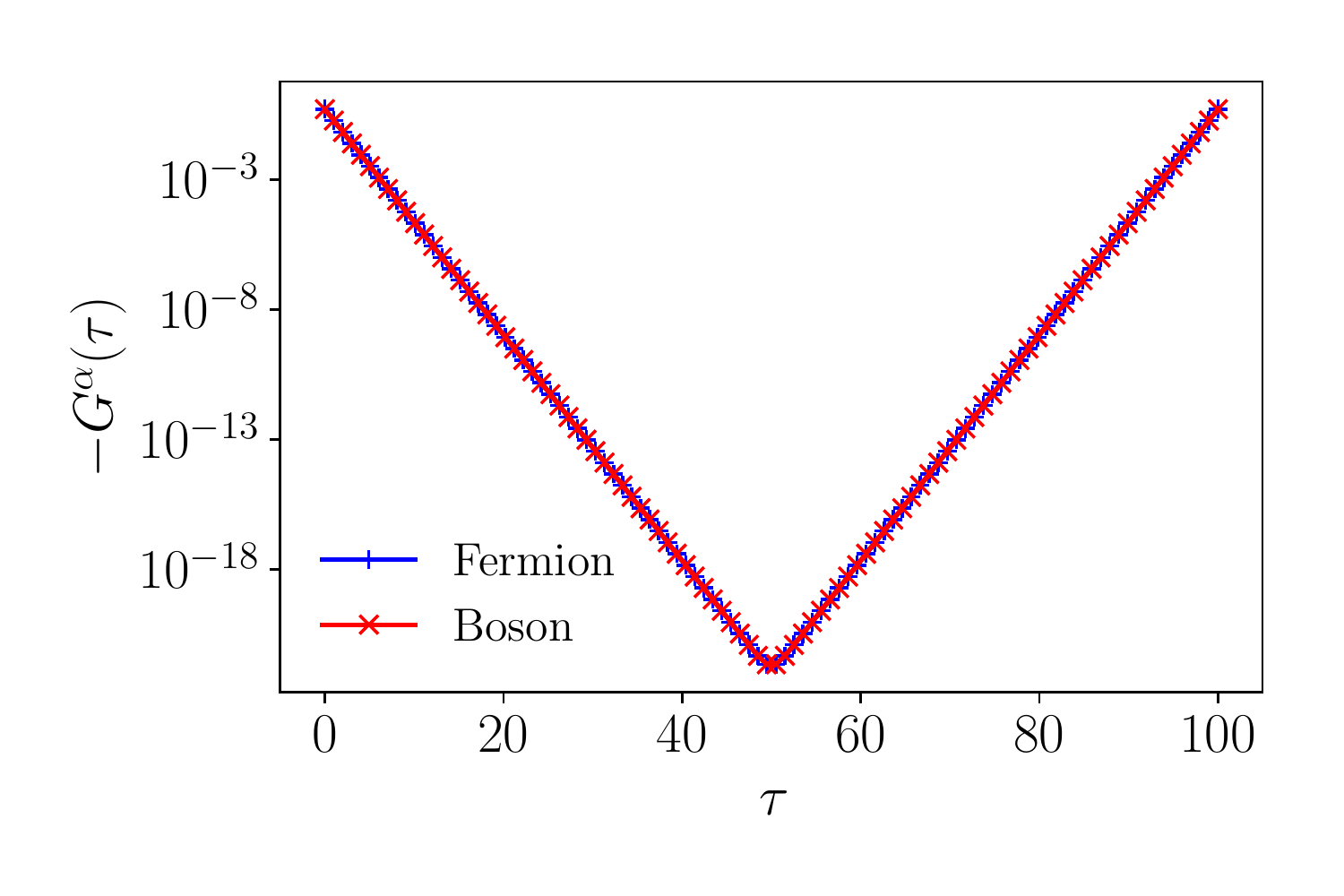}

	\begin{flushleft}
	\hspace{4em}	(c) Particle-hole asymmetric insulating model
    \end{flushleft}
     \centering
     \hspace{0.5cm}\includegraphics[width=0.3\textwidth,clip]{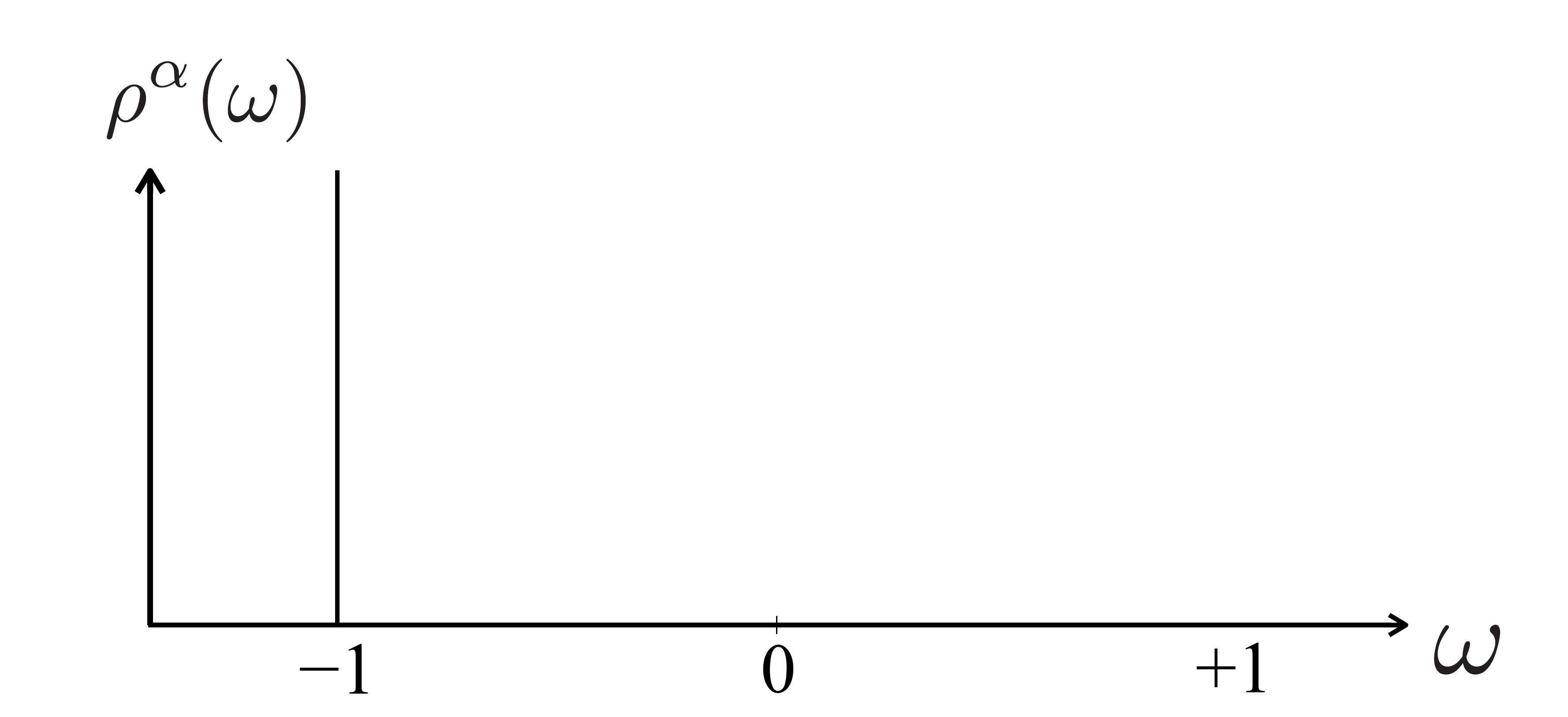}
     \includegraphics[width=0.35\textwidth,clip]{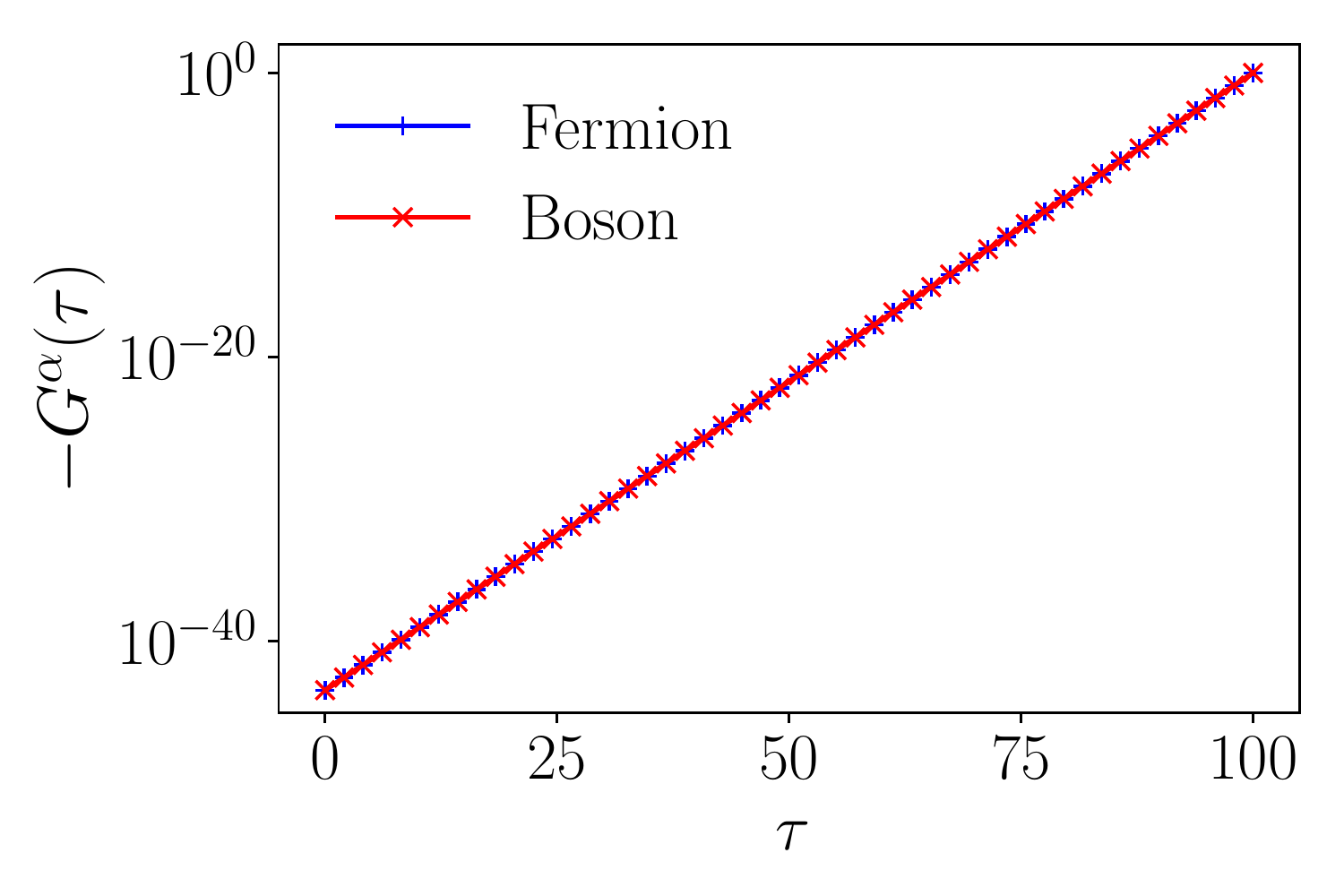}	
	
	\caption{
		(Color online)
		Spectral functions $\rho^{\alpha}(\omega)$ of (a) the gapless model, (b) the insulating model, and (c) the particle-hole asymmetric insulating model.
		Also plotted are the imaginary-time Green's function $G^\alpha(\tau)$ computed for $\beta=100$.
	}
	\label{fig:spectrum}
\end{figure}
This section describes how we assess the performance of the orthogonal representations.
The $\tau$-dependence of $G^{\alpha}(\tau)$ qualitatively differs between metals and insulators. 
We therefore consider three models represented by the spectra $\rho^\alpha(\omega)$ illustrated in Fig.~\ref{fig:spectrum}.
We refer to these models as ``gapless model", ``insulating model", and ``particle-hole asymmetric insulating model", respectively.
We use the energy range of $\rho^{\alpha}(\omega)$ as the unit of energy, i.e., $\Omega=1$.

The spectral function of the gapless model reads
\begin{align}
	\rho^\alpha(\omega) &=
     \begin{cases}
     	\frac{\pi}{2}\sqrt{1-\omega^2} & (\alpha = \mathrm{F})\\
     	\frac{3}{2}\sqrt{1-\omega^2} & (\alpha = \mathrm{B})
     \end{cases}.\label{eq:rho-model}
\end{align}
The physical spectrum for bosons is normalized as $\int_{-\infty}^\infty |\omega|\rho^\mathrm{B}(\omega) d\omega = 1$.
In Fig.~\ref{fig:spectrum}(a), we also show $G^\alpha(\tau)$ computed for $\beta = 100$.
We find that $G^\alpha(\tau)$ exhibits a power-law dependence for the gapless model rather than an exponential one around $\tau=0$ and $\beta$.
We also point out that $G^\mathrm{B}(\tau)$ decays faster than $G^\mathrm{F}(\tau)$.
This is because the physical spectrum for bosons $\omega \rho^\mathrm{B}
(\omega)$ vanishes linearly toward $\omega=0$.

Figure~\ref{fig:spectrum}(b) shows the spectral function of the insulating model, consisting of two delta peaks at $\omega = \pm 1$ as
\begin{align}
	\rho^\alpha(\omega) &= \frac{1}{2}\left\{\delta(\omega-1)+\delta(\omega+1)\right\}.
\end{align}
This resembles the spectrum of the atomic limit of correlated electrons, e.g., a Mott insulator.
For both fermions and bosons,
$G^\alpha(\tau)$ decays exponentially.
The spectral function of the particle-hole asymmetric insulating model is
given by
\begin{align}
	\rho^\alpha(\omega) &= \delta(\omega+1).
\end{align}

\begin{figure}
	\centering
	\includegraphics[width=0.95\linewidth,clip]{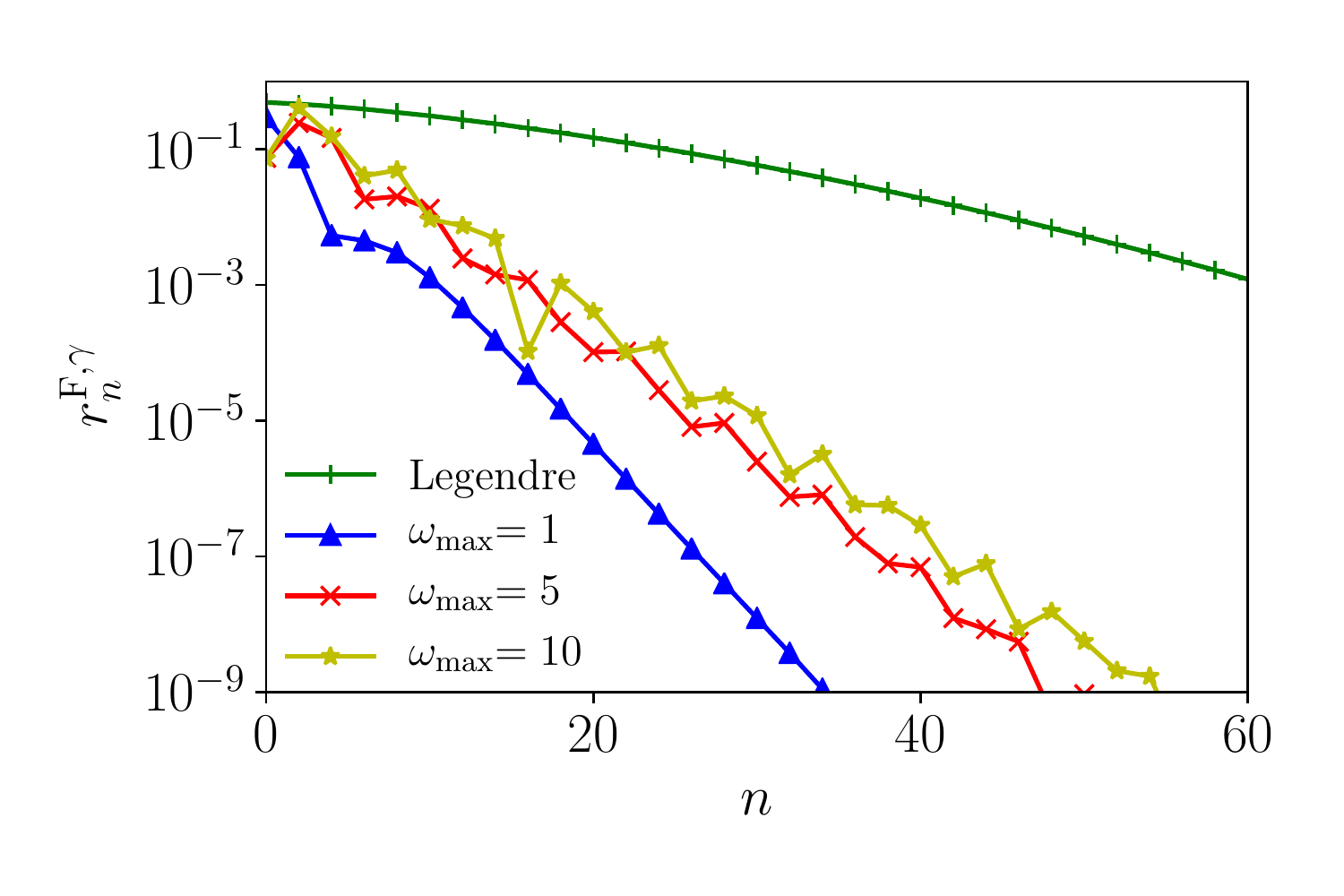}
	\caption{
		(Color online) The residual errors $r^{\alpha, R}_{n}$ computed for the fermionic gapless model at $\beta=1000$.
	}
	\label{fig:residual}
\end{figure}

We compute the expansion coefficients $G^{\alpha,\gamma}_l$ by Eq.~(\ref{eq:IR-decomp}) for the IR basis ($\gamma=\mathrm{IR}$) and by Eq.~(\ref{eq:LG-exp-coeff}) for the Legendre basis ($\gamma=\mathrm{L}$).
Using $G^{\alpha,\gamma}_l$ up to the $n$-th degree, we approximate $G^\alpha(\tau)$ as
\begin{align}
G^{\alpha,\mathrm{IR}}_n(\tau) &\equiv \sum_{l=0}^n G_l^{\alpha,\mathrm{IR}} U^\alpha_l(\tau),\\
G^{\alpha,\mathrm{L}}_n(\tau) &\equiv \frac{1}{\beta} \sum_{l=0}^n\sqrt{2l+1}  G_l^{\alpha,\mathrm{L}} P_l(x(\tau)).
\end{align}
We then define the residual of the expansion by
\begin{align}
r^{\alpha, \gamma}_{n}
& \equiv \underset{0 \le \tau \le \beta}{\mathrm{max}} |G^\alpha(\tau) - G^{\alpha, \gamma}_n(\tau)|.
\end{align}
Figure~\ref{fig:residual} shows a typical $n$ dependence of $r^{\alpha, \gamma}_{n}$ at low temperatures.
Although $r^{\alpha, \gamma}_{n}$ decreases exponentially for both IR and Legendre bases, 
its exponent is much smaller (faster decay) for the IR basis than the Legendre basis.
We now quantify the performance of the two basis sets using the minimum number of basis functions, $\Nag$, such that $r^{\alpha, \gamma}_{n} < 10^{-5}$ for $\forall n \ge \Nag$.
For the data in Fig.~\ref{fig:residual}, we obtained $\NFIR=20$, $24$, $30$ for the IR basis ($\Lambda=1$, $5$, $10$, respectively) and $\NFL=88$ for the Legendre basis.
Note that expansion coefficients for odd $l$'s vanish
since the model spectrum has the particle-hole symmetry.

\subsection{Results}\label{sec:model-analysis-results}
\begin{figure}
	\begin{flushleft}\hspace{4em}(a) Gapless model\end{flushleft}\vspace{-1em}
	\centering
	\includegraphics[width=0.45\textwidth,clip]{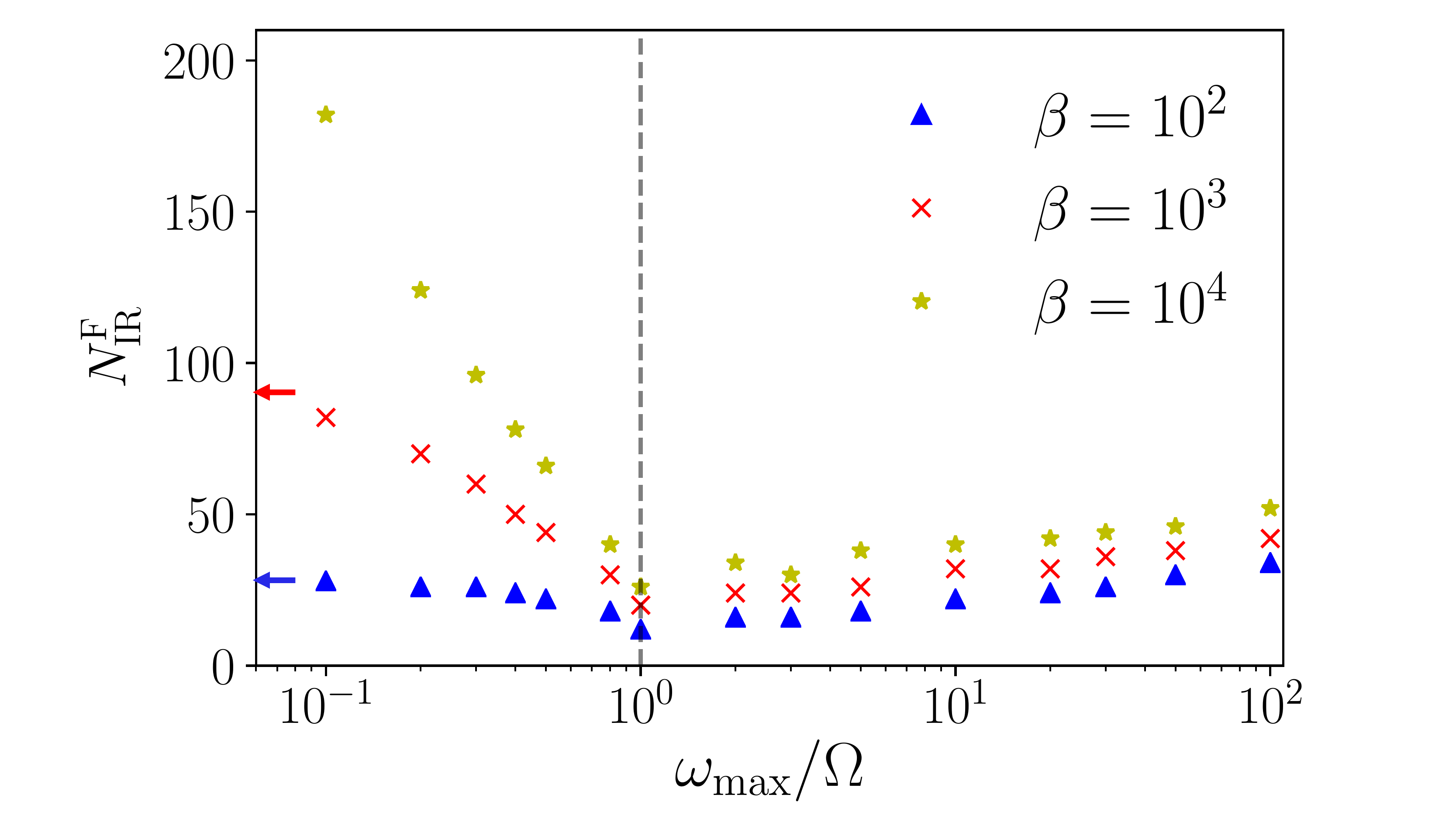}
	
	\begin{flushleft}\hspace{4em}(b) Insulating model \end{flushleft}\vspace{-1em}
	\centering
	\includegraphics[width=0.45\textwidth,clip]{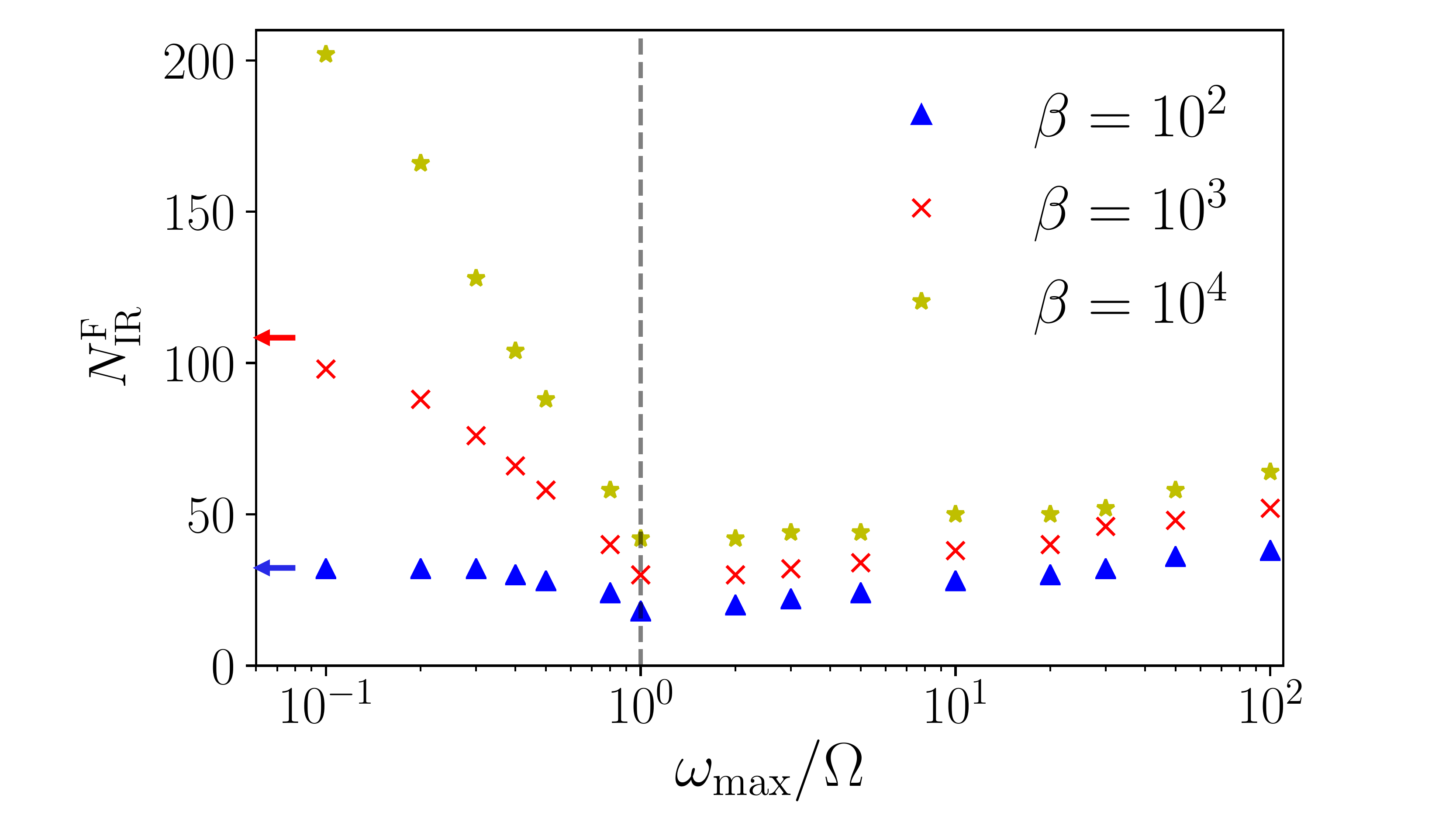}
	
	\begin{flushleft}\hspace{4em}(c) Particle-hole asymmetric insulating model \end{flushleft}\vspace{-1em}
	\centering
	\includegraphics[width=0.45\textwidth,clip]{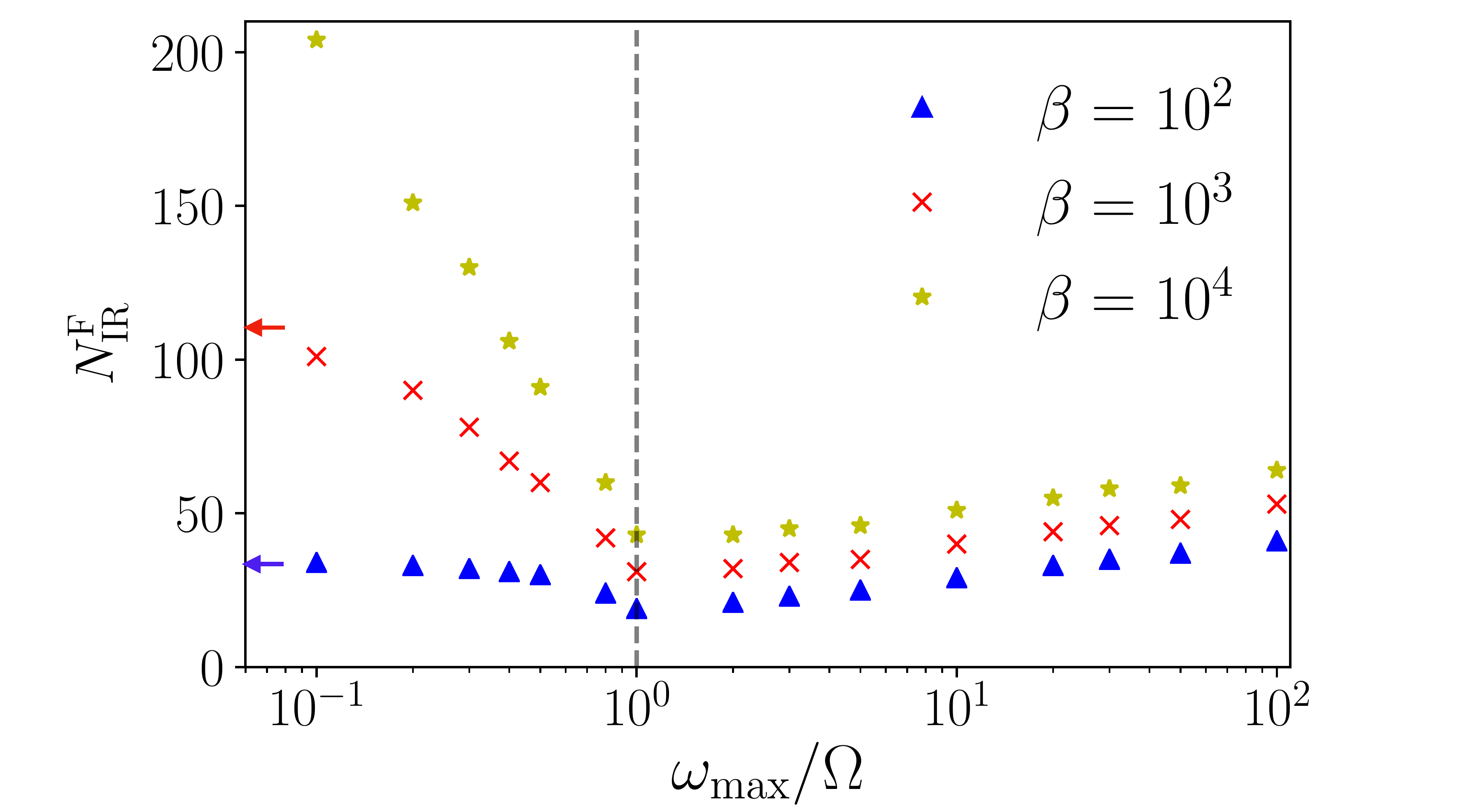}
	
	\caption{
		(Color online) The minimum number of basis functions, $\NFIR$, as a function of $\wmax/\Omega$ for fixed $\beta$. (a) The gapless model, (b) insulating model and (c) particle-hole asymmetric insulating model. The expansion is physical for $\wmax/\Omega \ge 1$.
		The arrows represent the results for the Legendre representation.
	}
	\label{fig:Lambda-dep}
\end{figure}

We now analyze the performance of the IR and Legendre representations based on the quantity $\Nag$ introduced in the last subsection.
Since the IR basis has the control parameter $\Lambda \equiv \beta\wmax$,
we first clarify how the performance changes as $\Lambda$ is varied.
Figure~\ref{fig:Lambda-dep} shows $\NFIR$ as a function of $\wmax$ for fixed values of $\beta$.
Both the gapless and insulating models exhibit a similar feature:
$\NFIR$ takes the minimum at $\wmax/\Omega=1$, and increases only logarithmically as $\wmax$ is increased.
For $\wmax/\Omega < 1$, on the other hand,
$\NFIR$ grows very rapidly as $\wmax$ is decreased.
This behavior is consistent with the fact that the fast convergence of the expansion is ensured only for $\wmax/\Omega \ge1$ [see Eq.~(\ref{eq:gl})].
Note that the limit of $\wmax = 0$ is equivalent to the Legendre representation.
In practical calculations, one may set $\wmax$ to a value much larger than a rough estimate of $\Omega$.
As shown above, this only slightly influences the performance.
For boson, $\NBIR$ shows the same behavior as above (not shown).
\begin{figure}
	\begin{flushleft}\hspace{4em}(a) Gapless model\end{flushleft}\vspace{-1em}
	\centering
	\includegraphics[width=0.42\textwidth,clip]{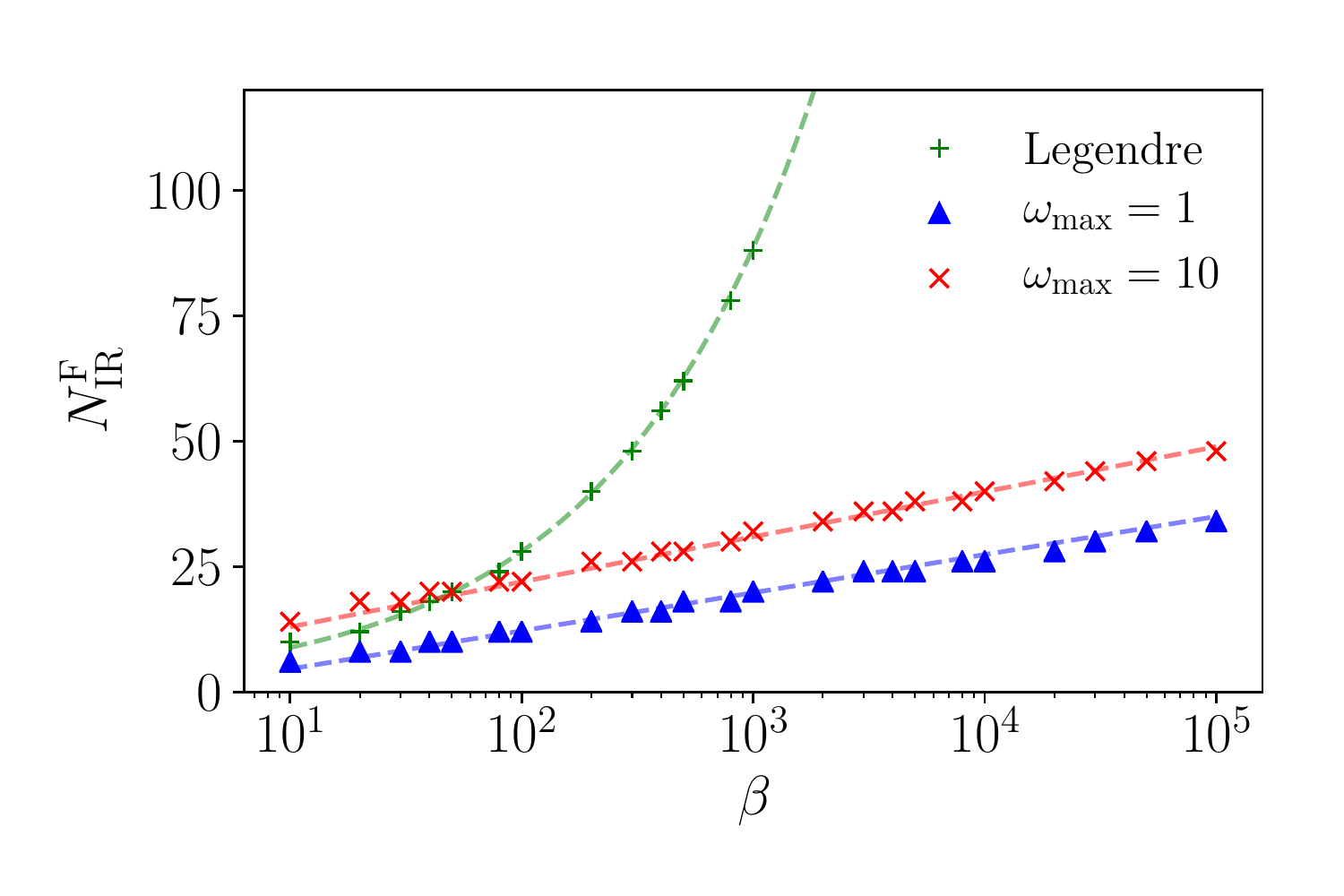}
	\includegraphics[width=0.42\textwidth,clip]{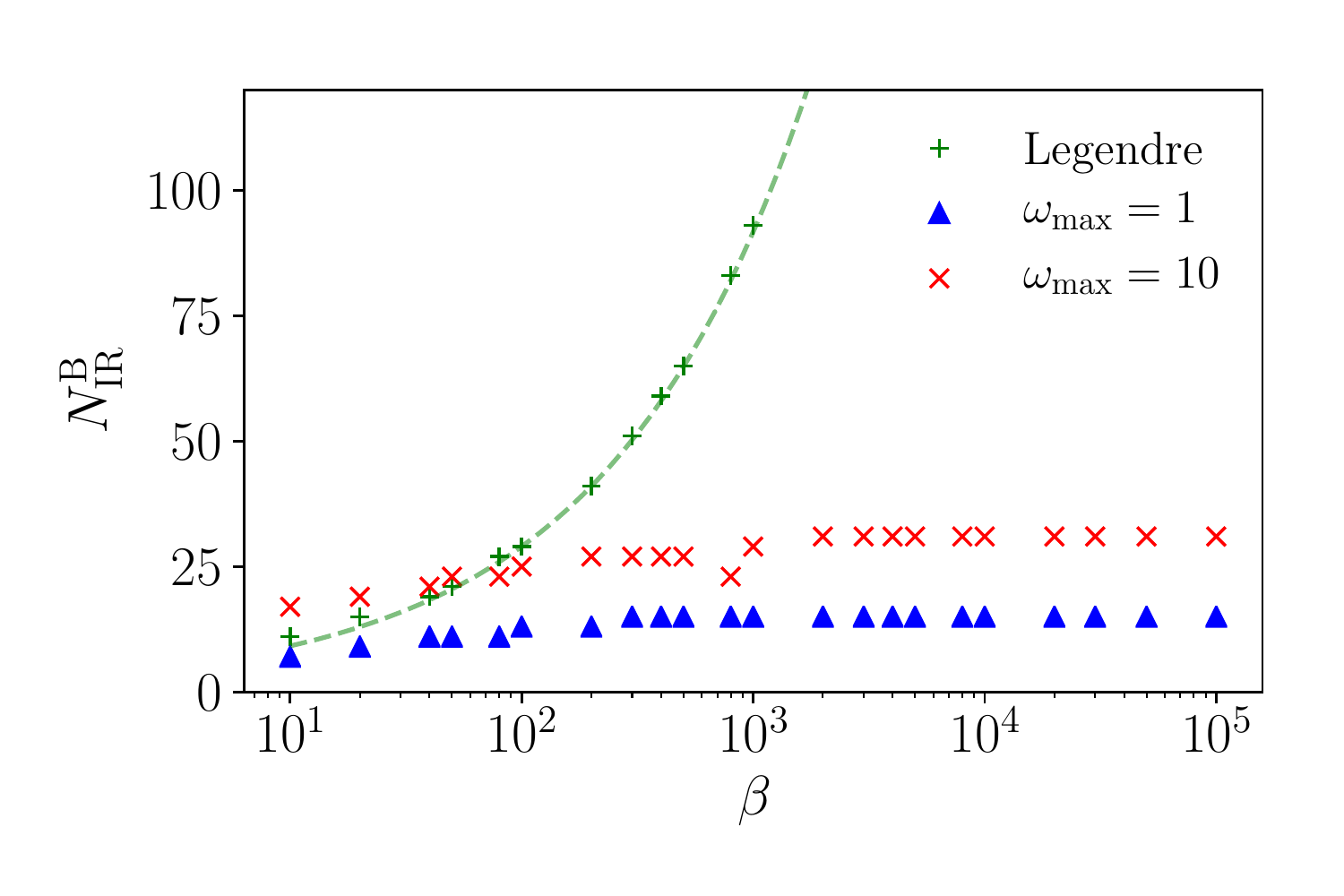}
	\begin{flushleft}\hspace{4em}(b) Insulating model \end{flushleft}\vspace{-1em}
	\centering
	\includegraphics[width=0.42\textwidth,clip]{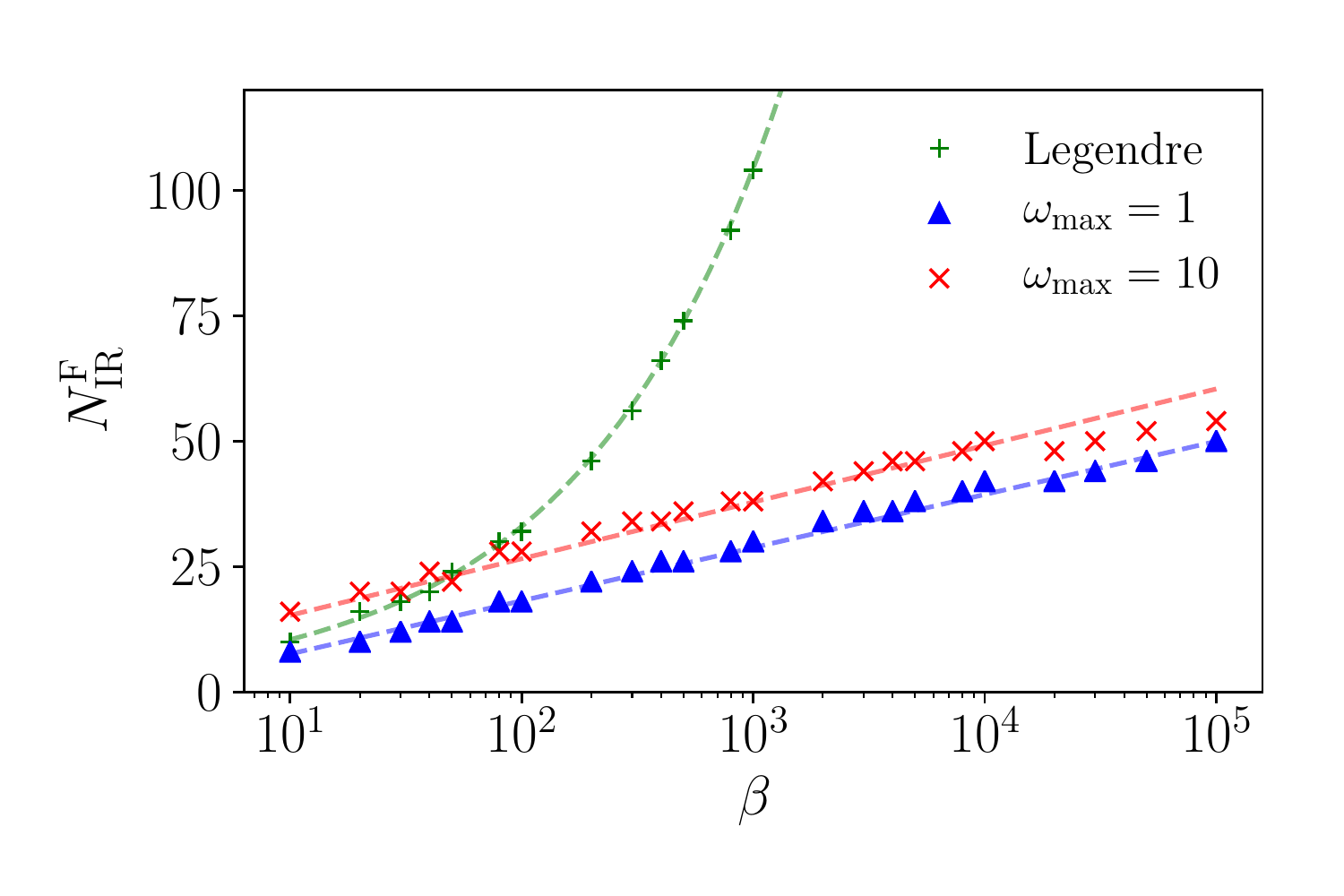}
	\includegraphics[width=0.42\textwidth,clip]{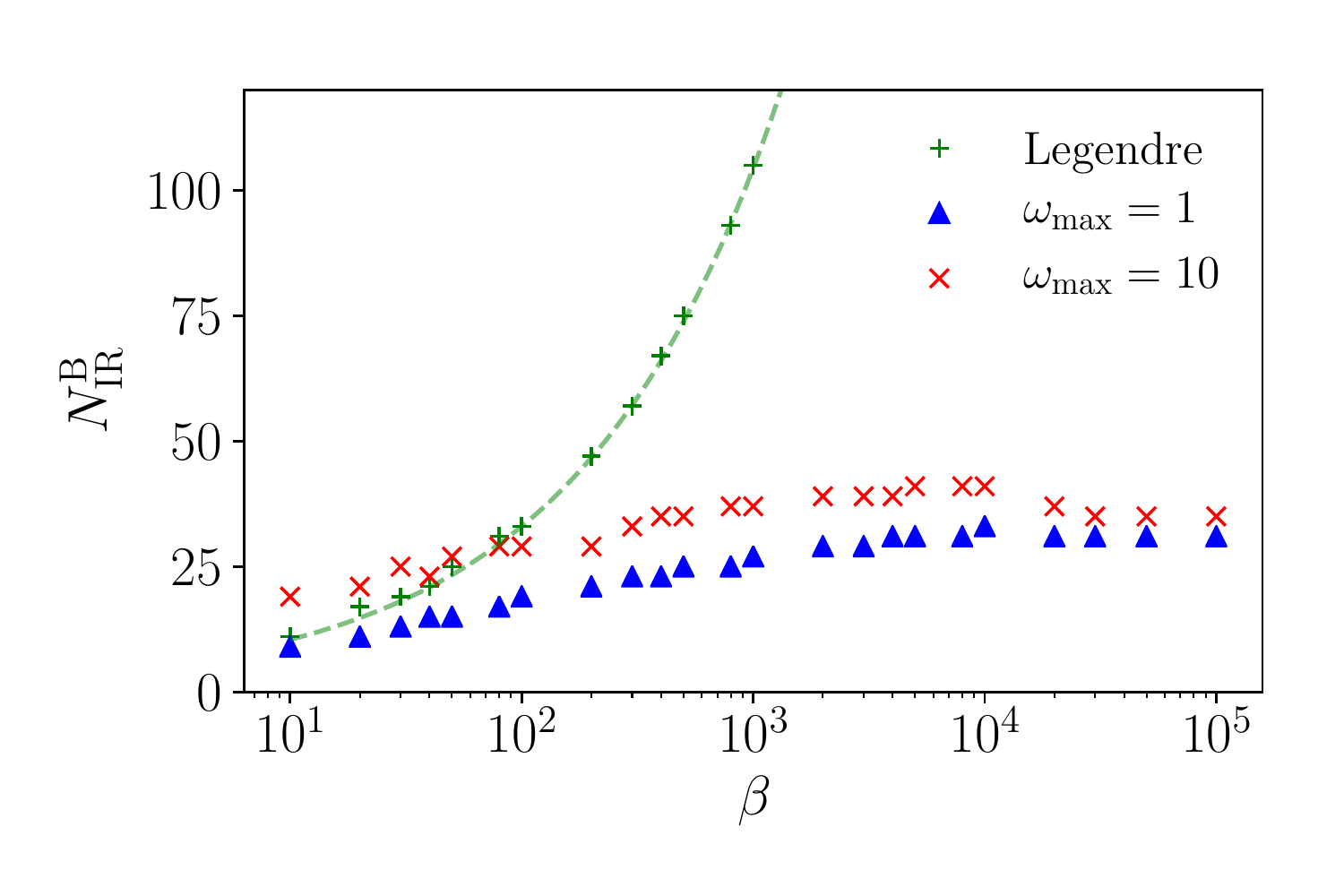}
	\caption{
		(Color online) Number of the required basis functions, $\Nag$, as a function of $\beta$.
		(a) gapless model and (b) insulating model.
		The broken curve ($\propto \beta^{0.5}$) and lines ($\propto \log \beta$) are used to guide the eye.
	}
	\label{fig:scaling}
\end{figure}

We next discuss the $\beta$ dependence of $\Nag$ for fixed $\wmax$ (1 and 10).
From now on, we present results only for the gapless model and the insulating model since we obtained qualitatively the same results for the particle-hole asymmetric model.
Figure~\ref{fig:scaling}(a) shows the results obtained for the gapless model.
For the Legendre basis,
$\NaL$ grows rapidly as $\beta$ is increased.
We found that this increase follows a power law $\NaL \propto \beta^{0.5}$ as indicated by the broken curve.
This power law is understood from the asymptotic zero distribution of $P_l(x)$ at large $l$.
We refer the interested reader to Appendix~\ref{appendix:legendre} for more details.

For the IR, on the other hand, $\NaIR$ grows much more slowly.
For fermions, $\NFIR$ follows $\NaIR \propto \log \beta$ as indicated by lines in Fig.~\ref{fig:scaling}(a).
The IR becomes superior to the Legendre basis for $\beta \gtrsim 10$ -- $100$.
Remarkably, for bosons, $\NBIR$ even converges to a constant at large $\beta$.
This indicates that only a fixed number of basis functions are sufficient for describing $G^\mathrm{B}(\tau)$ within a given tolerance at low temperature.

Figure~\ref{fig:scaling}(b) shows the results obtained for the insulating model.
For both of the Legendre basis and the IR basis,
we need slightly more basis functions than the gapless model to reach convergence to the same tolerance.
This is because $G^\alpha(\tau)$ of the insulating model has a stronger $\tau$ dependence than the gapless model around $\tau=0$ and $\beta$ (see Fig.~\ref{fig:spectrum}).
We however observed qualitatively the same $\beta$ dependences as the gapless model: The number of basis functions grows logarithmically with $\beta$, and $\NBIR$ is smaller than $\NFIR$ at large $\beta$.
These results demonstrate the universality of these features of the IR.

Although the primary interest here is the convergence properties of $G^\alpha(\tau)$,
some of the readers may be interested in how well the spectral function $\rho^{\alpha}(\omega)$ is described by $\{V_l^\alpha(\omega)\}$ for $l \le \NaIR$.
As indicated by Eq.~(\ref{eq:gl}), $\rho^\alpha_l$ decays slower than $G_l^{\alpha,\mathrm{IR}}$.
We emphasize, however, that the fast convergence of $G^\alpha(\tau)$ is sufficient for performing efficient calculations of quantum many-body theories such as the Bethe-Salpeter equation for the susceptibility.
We refer the interested readers to Appendix~\ref{appendix:spectrum} and Ref.~\onlinecite{Otsuki:2017er} for results on the convergence of $\rho^{\alpha}(\omega)$.

\section{Analysis of IR basis functions}\label{sec:detail}
\begin{figure}
	\begin{flushleft}\hspace{2em}(a)\end{flushleft}	\centering\vspace{-1em}
	\includegraphics[width=0.42\textwidth,clip]{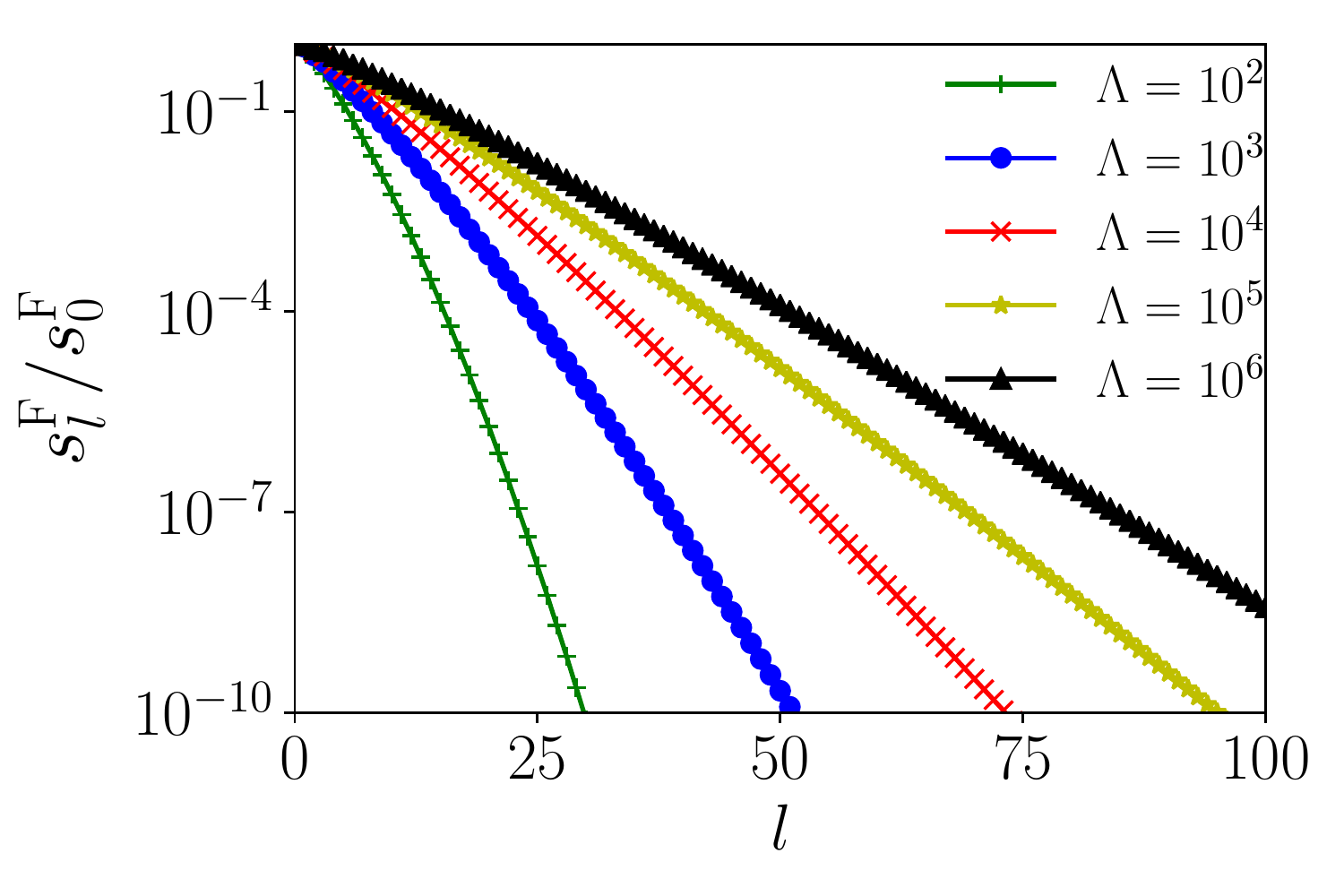}
	\begin{flushleft}\hspace{2em}(b)\end{flushleft}	\centering\vspace{-1em}
	\includegraphics[width=0.42\textwidth,clip]{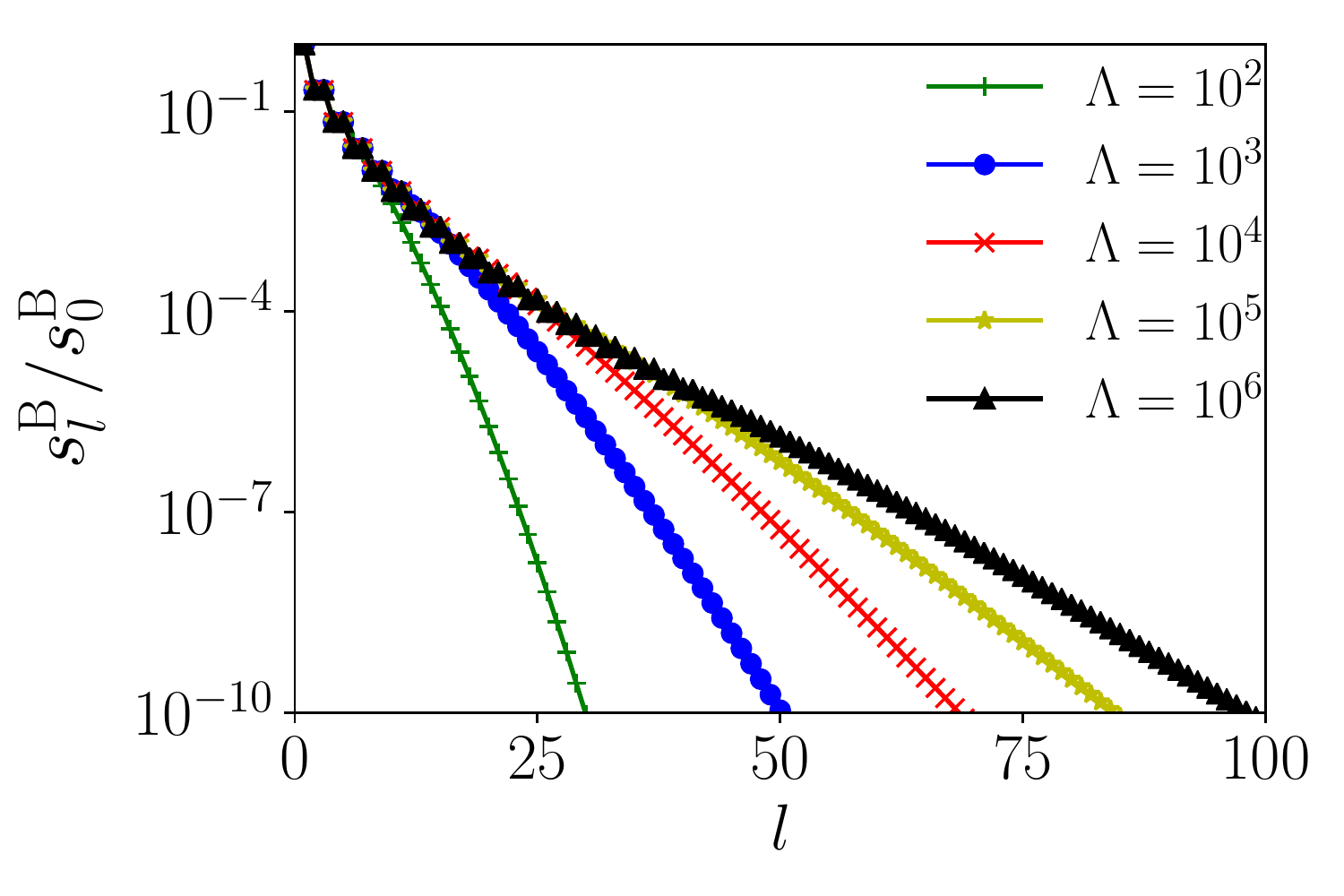}
	\caption{
		(Color online) Singular values $s_l^\alpha$ computed for (a)~$\alpha=\mathrm{F}$ and (b)~$\alpha=\mathrm{B}$.
	}
	\label{fig:sl}
\end{figure}

\begin{figure}
	\begin{flushleft}\hspace{2em}(a)\end{flushleft}	\centering\vspace{-1em}
	\includegraphics[width=0.42\textwidth,clip]{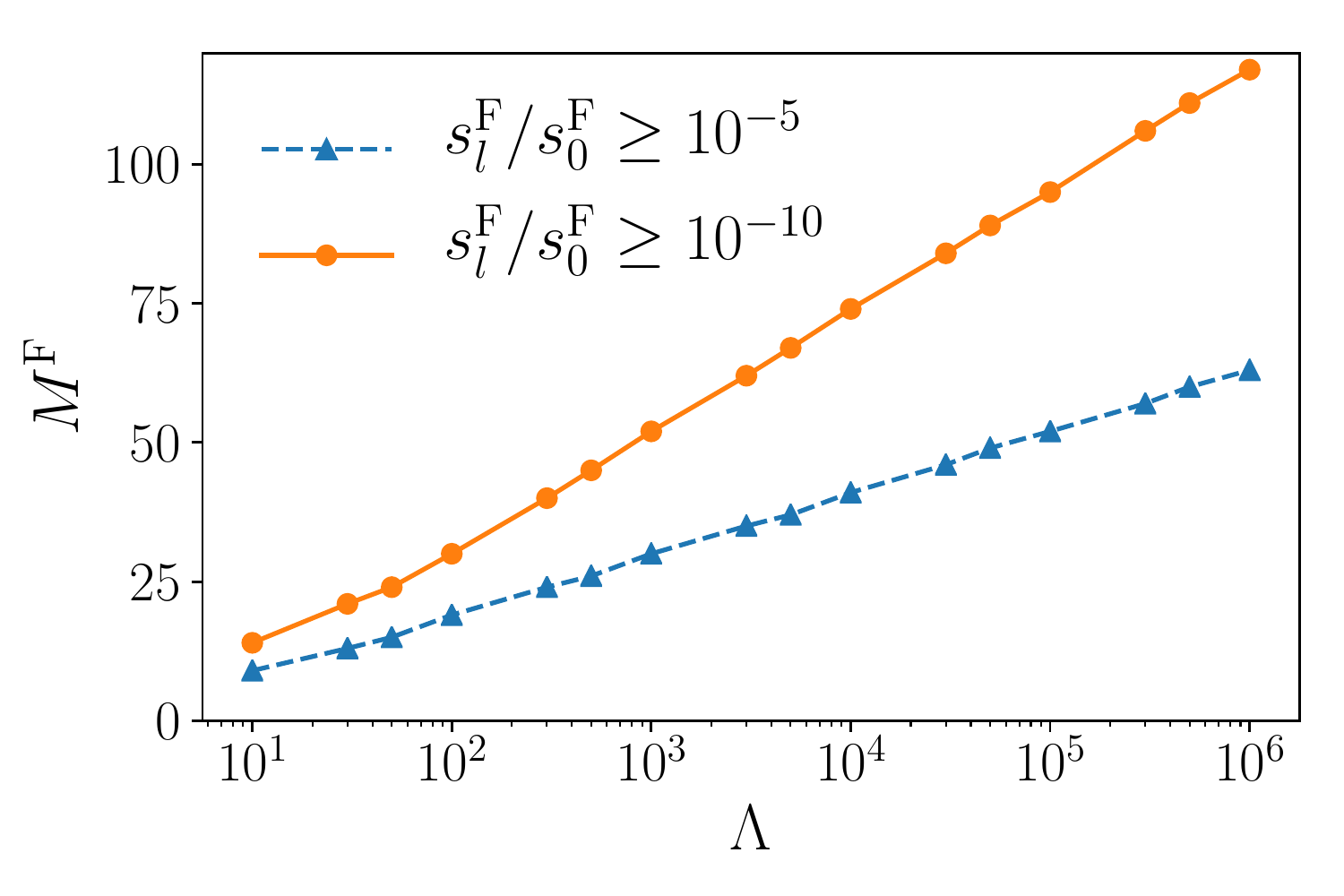}
	\begin{flushleft}\hspace{2em}(b)\end{flushleft}	\centering\vspace{-1em}
	\includegraphics[width=0.42\textwidth,clip]{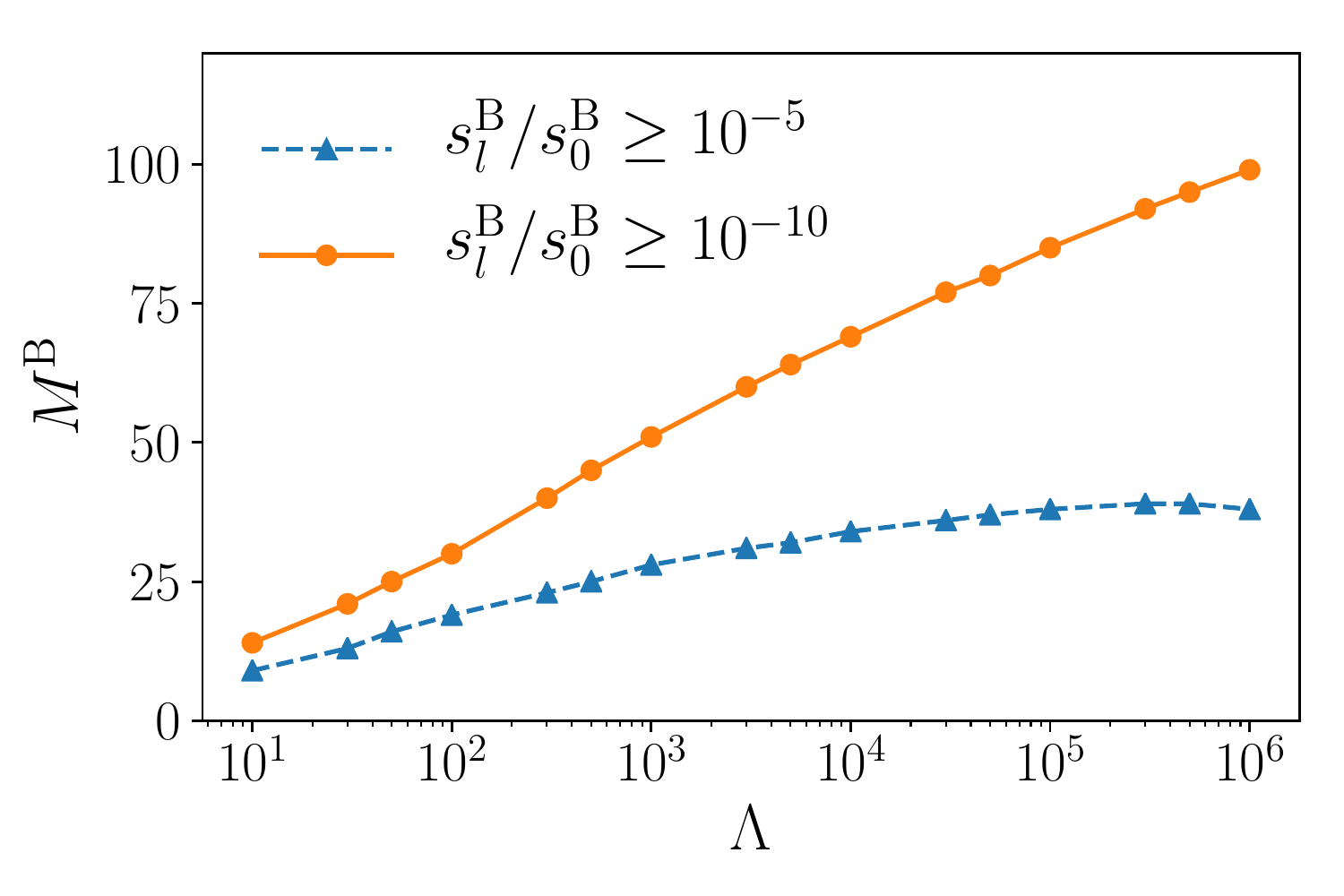}
	\caption{
		(Color online) Number of singular values above a fixed cut-off value for (a) fermion and (b) boson.
	}
	\label{fig:sl-cut}
\end{figure}
\begin{figure}
	\begin{flushleft}\hspace{4em}(a) Fermion\end{flushleft}	\centering
		\vspace{-1em}
	\includegraphics[width=0.3\textwidth,clip]{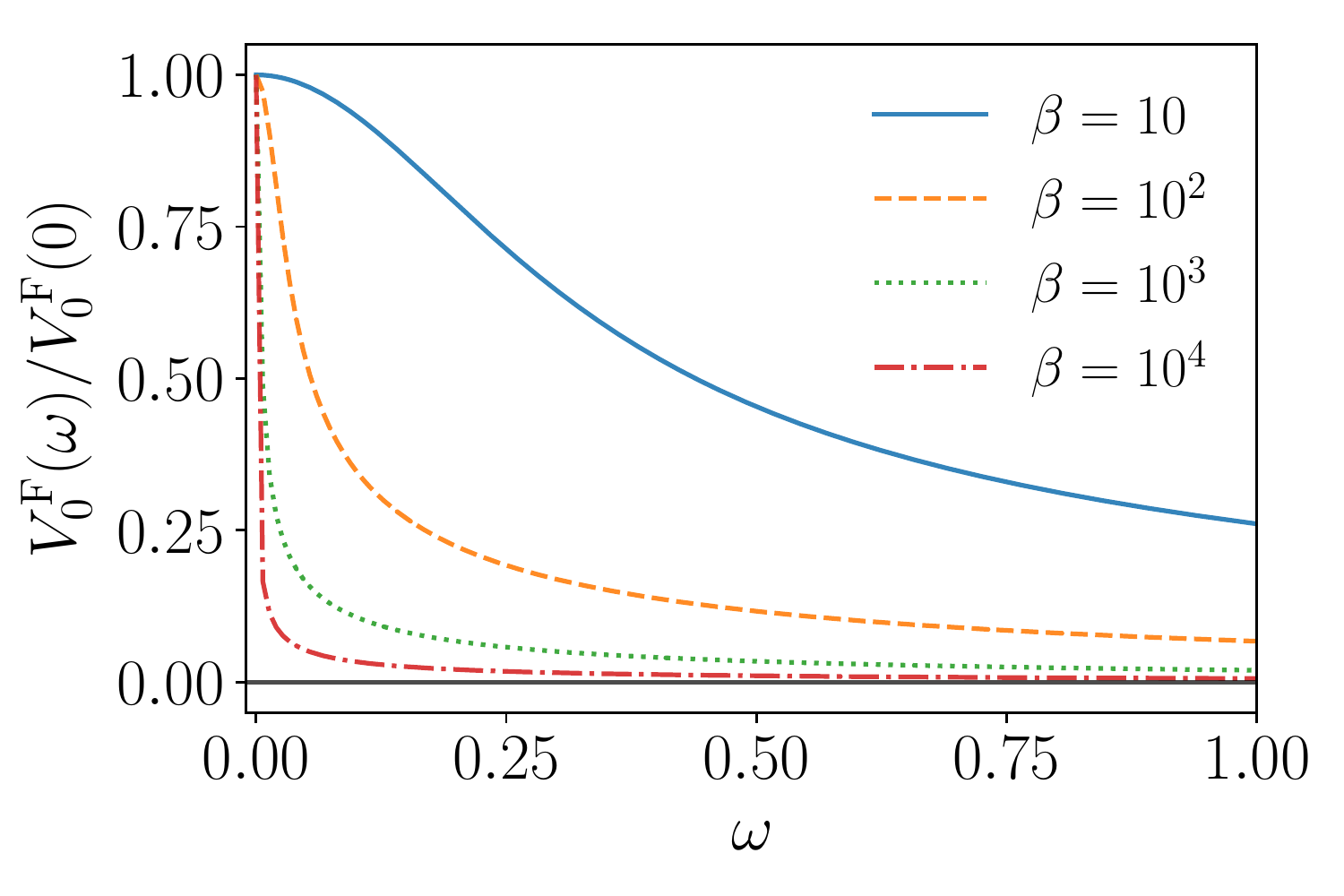}
	\includegraphics[width=0.3\textwidth,clip]{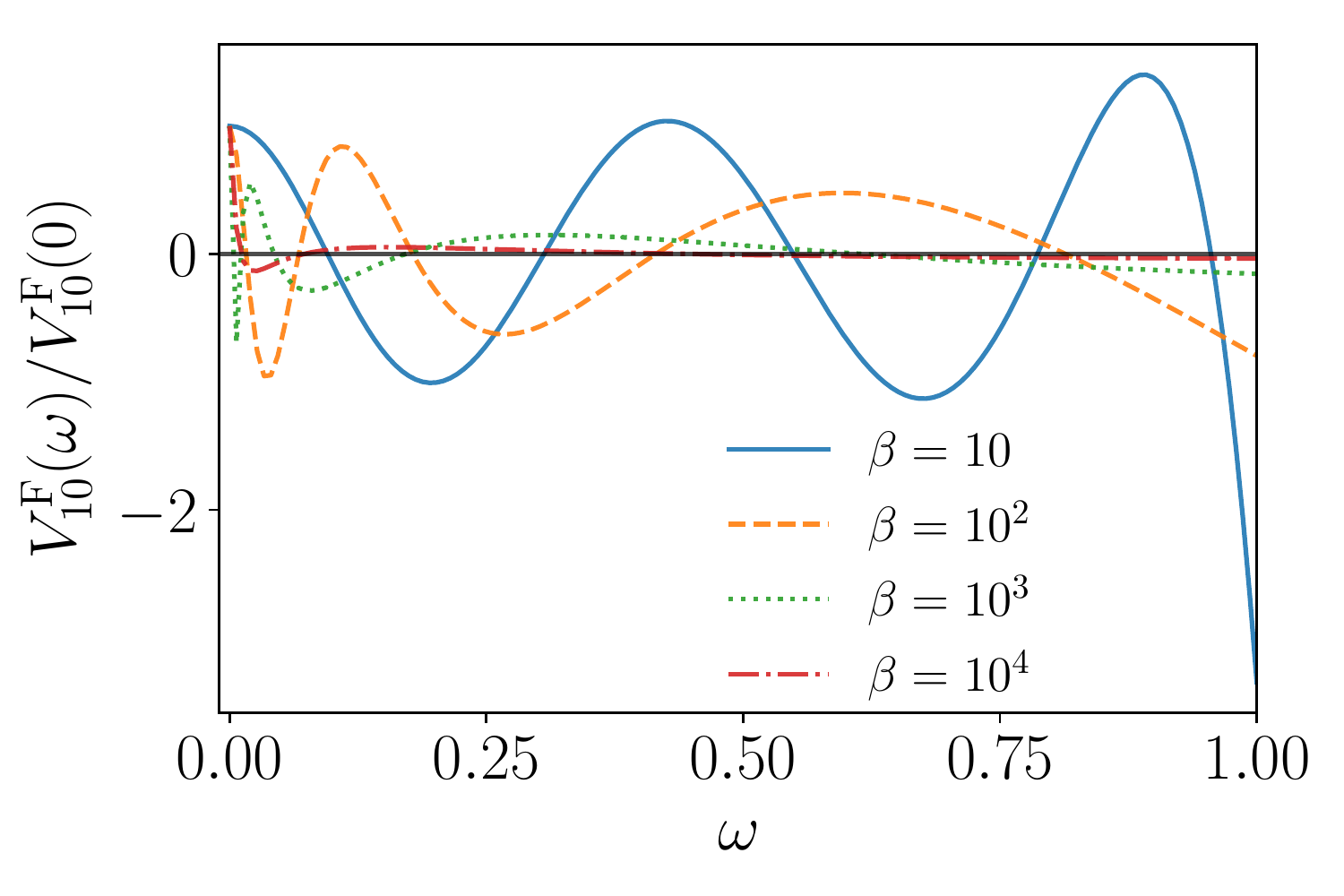}
	\includegraphics[width=0.3\textwidth,clip]{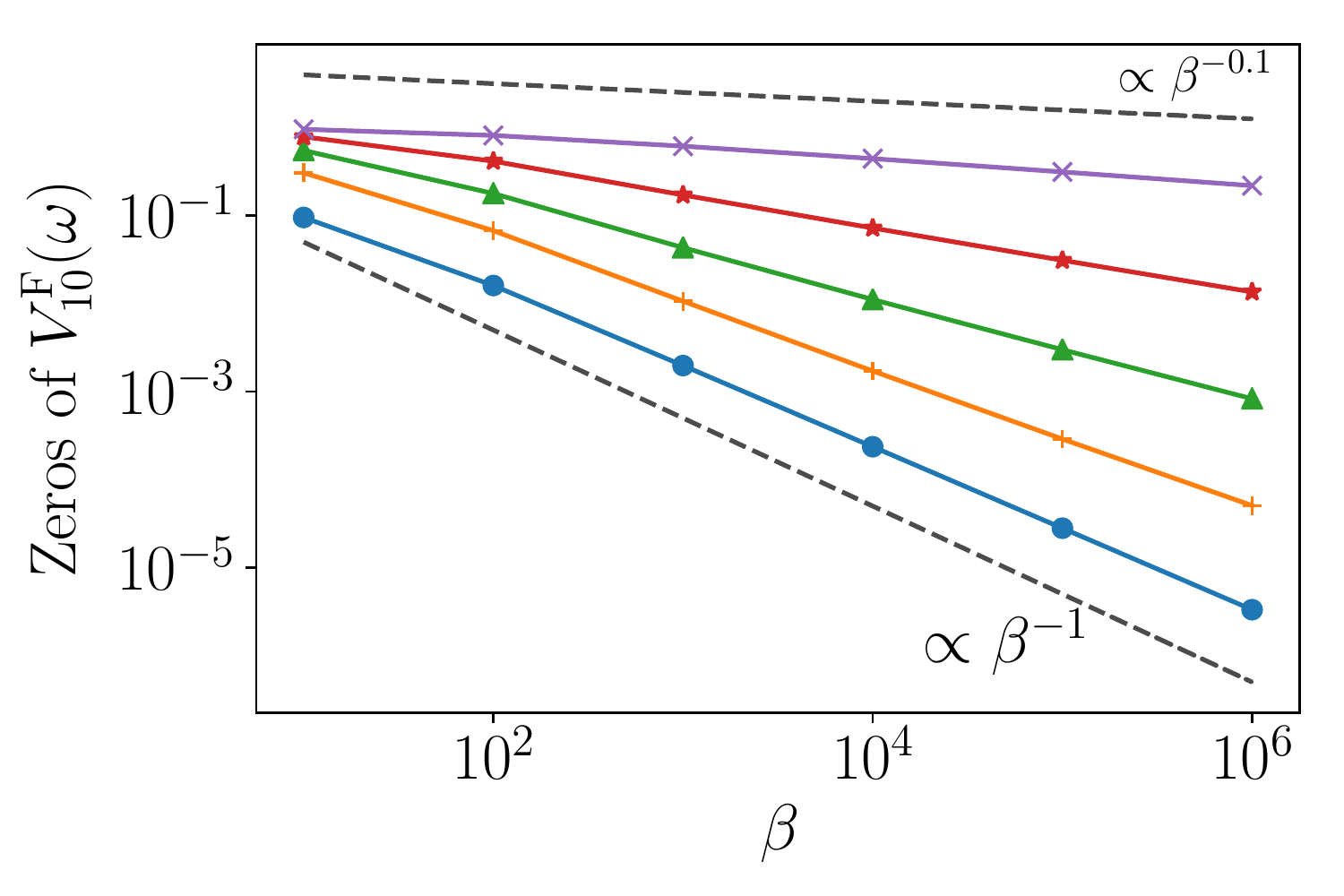}
		\vspace{-1em}
	\begin{flushleft}\hspace{4em}(b) Boson\end{flushleft}	\centering
		\vspace{-1em}
	\includegraphics[width=0.3\textwidth,clip]{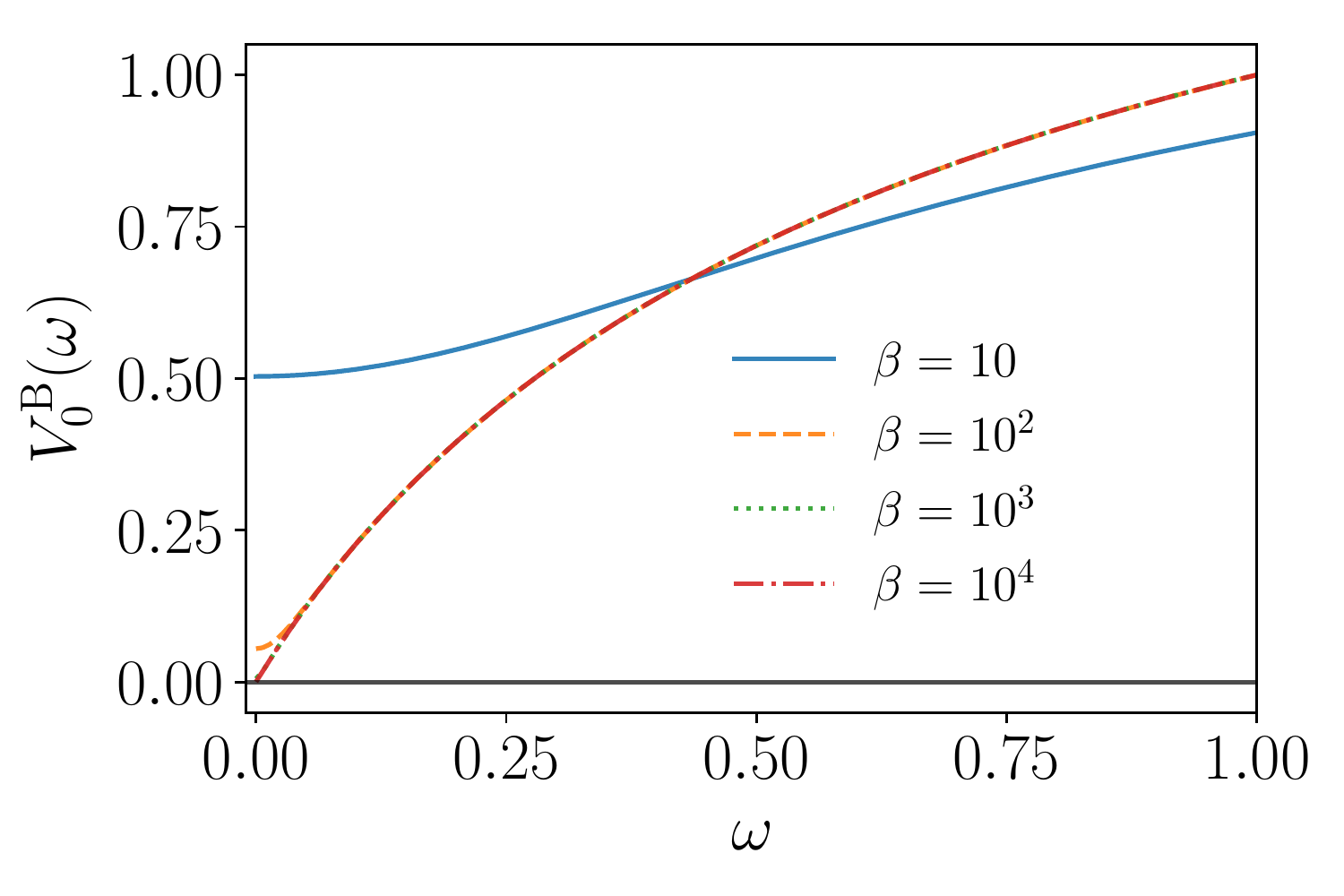}
	\includegraphics[width=0.3\textwidth,clip]{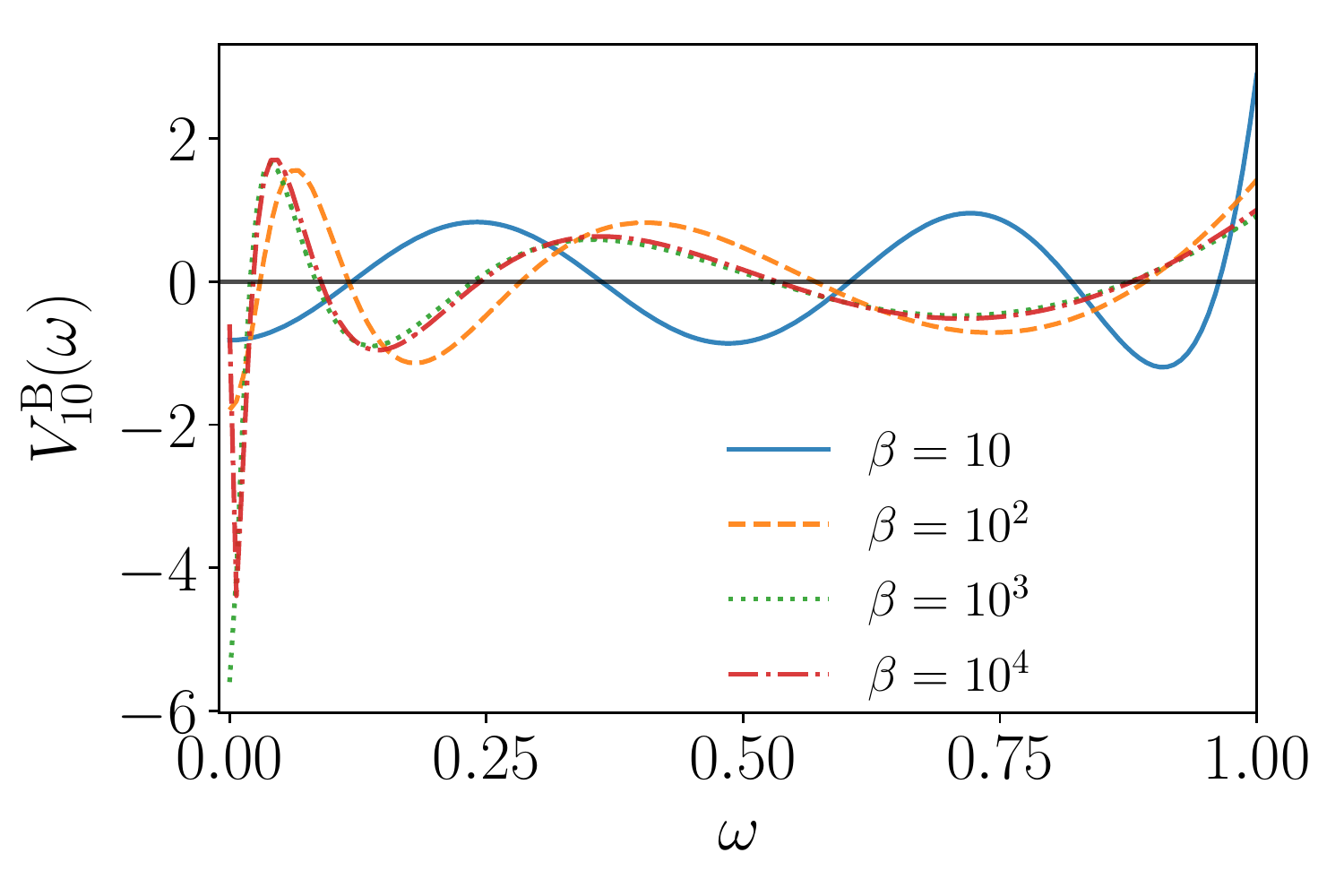}
	\includegraphics[width=0.3\textwidth,clip]{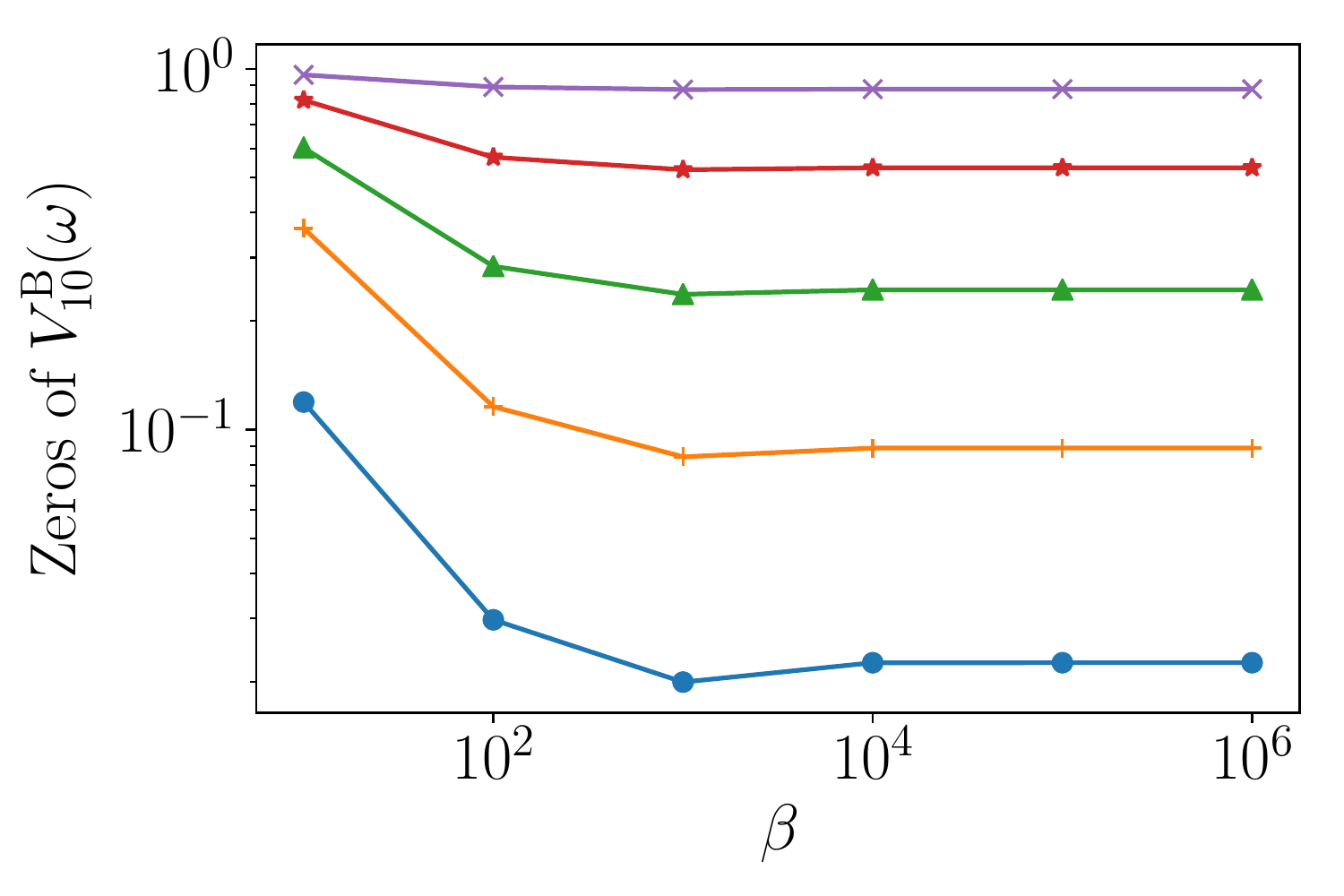}
	\caption{(Color online)
		$\beta$ dependence of $V_l^\alpha(\omega)$ for $l=0$ and 10 ($\wmax=1$). We also show the $\beta$ dependence of the positions of the zeros in the interval $[0,~\wmax]$.
			In the third figure of (a), the broken lines denote $\propto \beta^{-1}$ and $\propto \beta^{-0.1}$, respectively.
	}
	\label{fig:vly}
\end{figure}
\begin{figure}
	\begin{flushleft}\hspace{4em}(a) Fermion\end{flushleft}	\centering
	\vspace{-1em}
	\includegraphics[width=0.3\textwidth,clip]{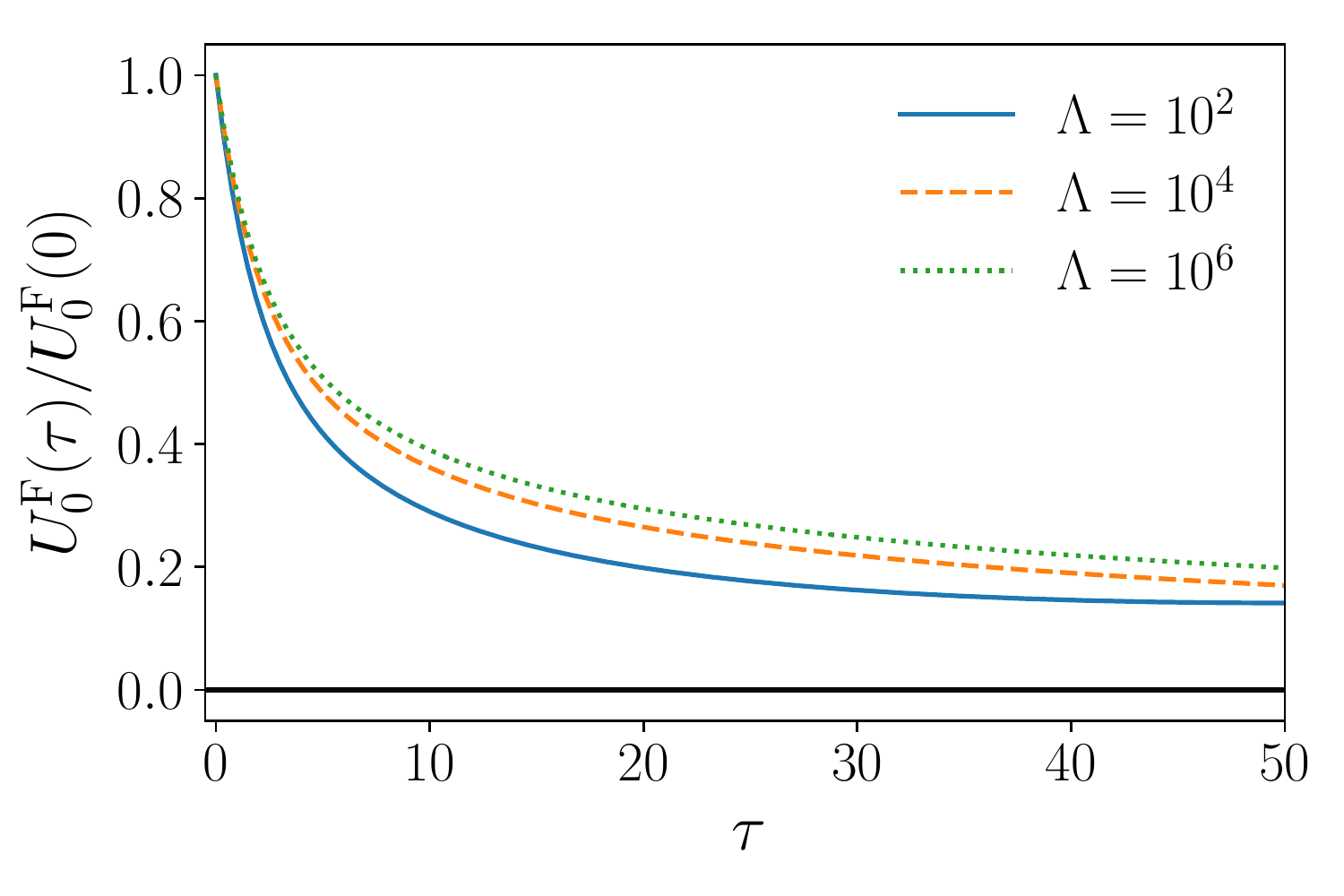}
	\includegraphics[width=0.3\textwidth,clip]{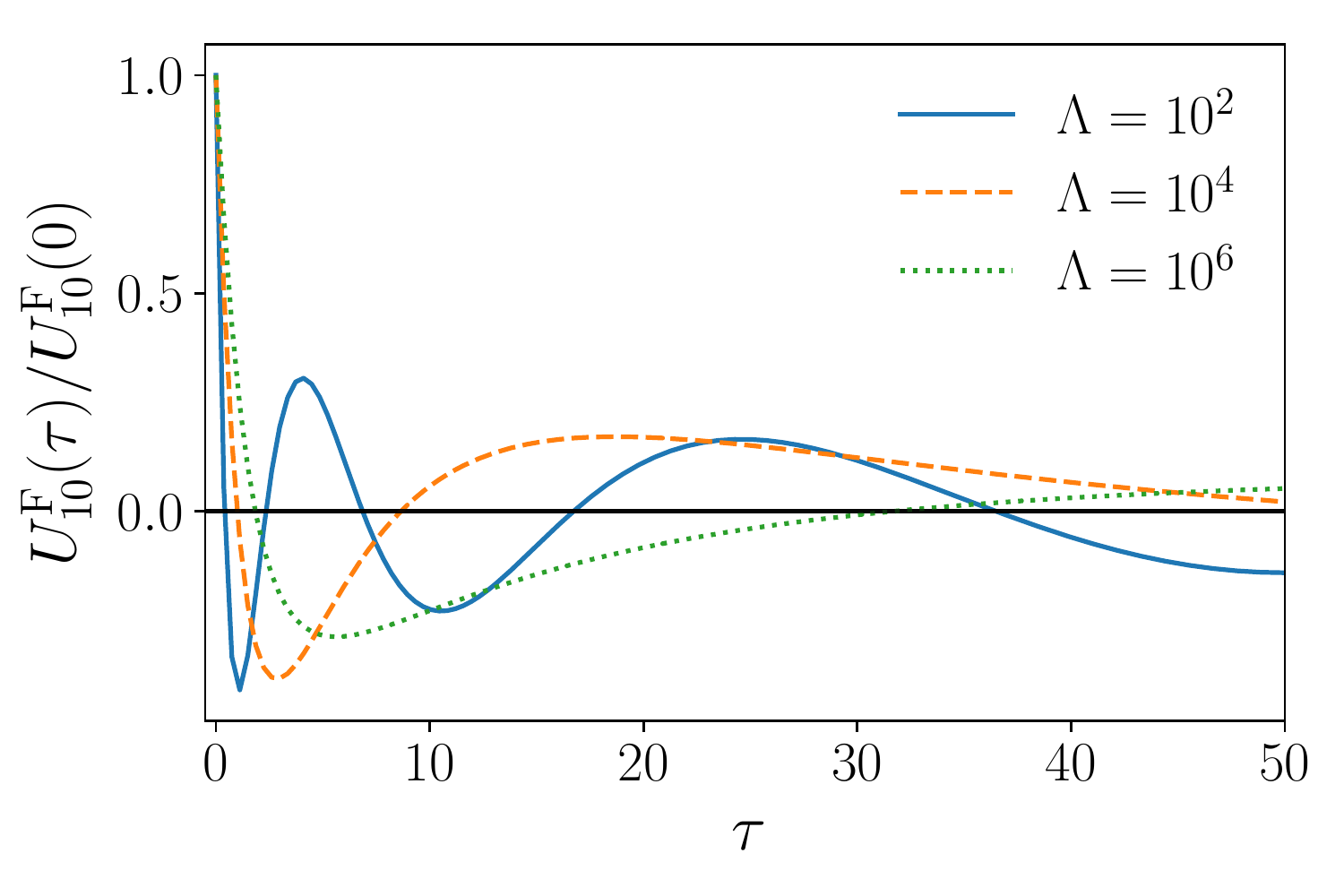}
	\includegraphics[width=0.3\textwidth,clip]{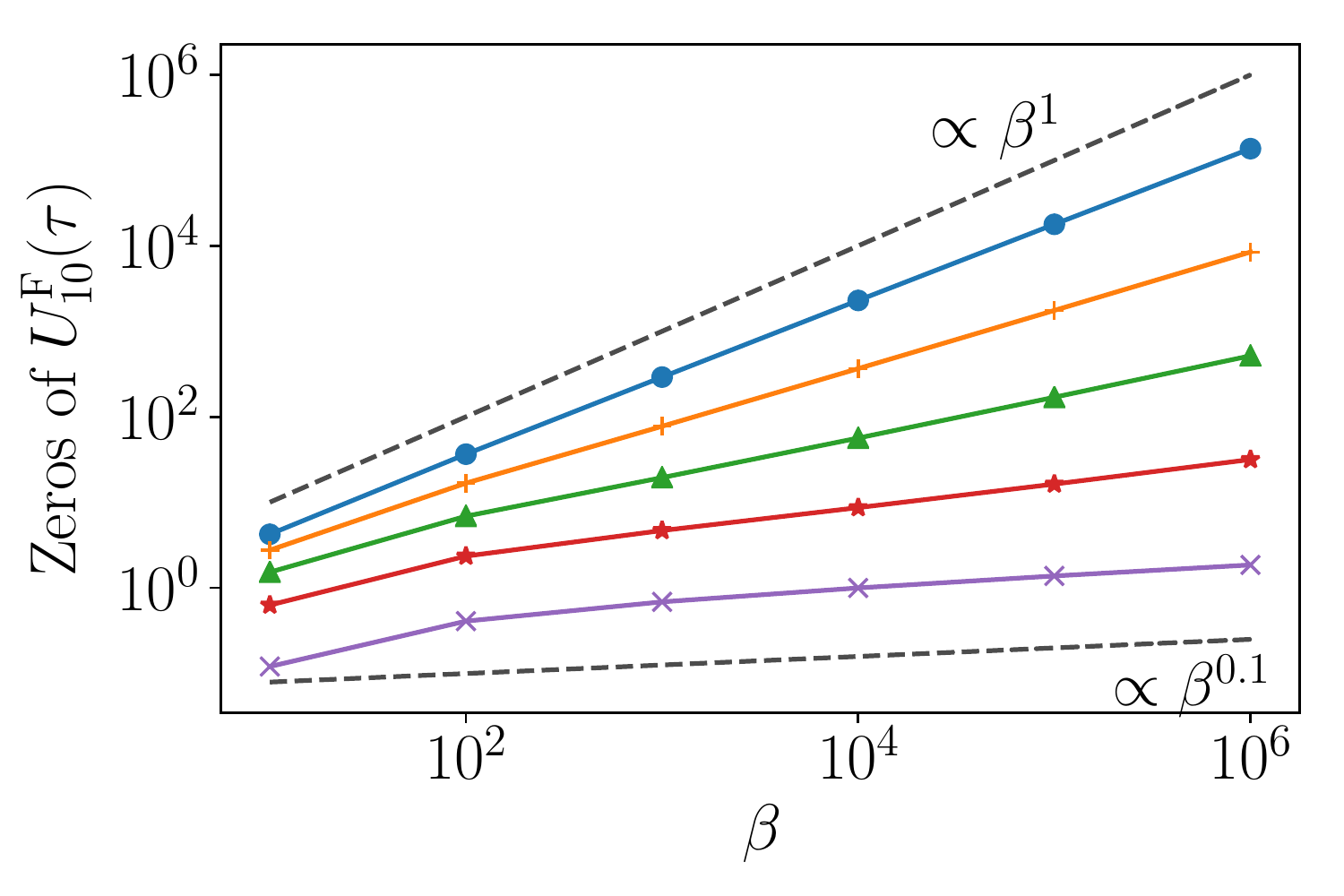}
		\vspace{-1em}
	\begin{flushleft}\hspace{4em}(b) Boson\end{flushleft}	\centering
		\vspace{-1em}
	\includegraphics[width=0.3\textwidth,clip]{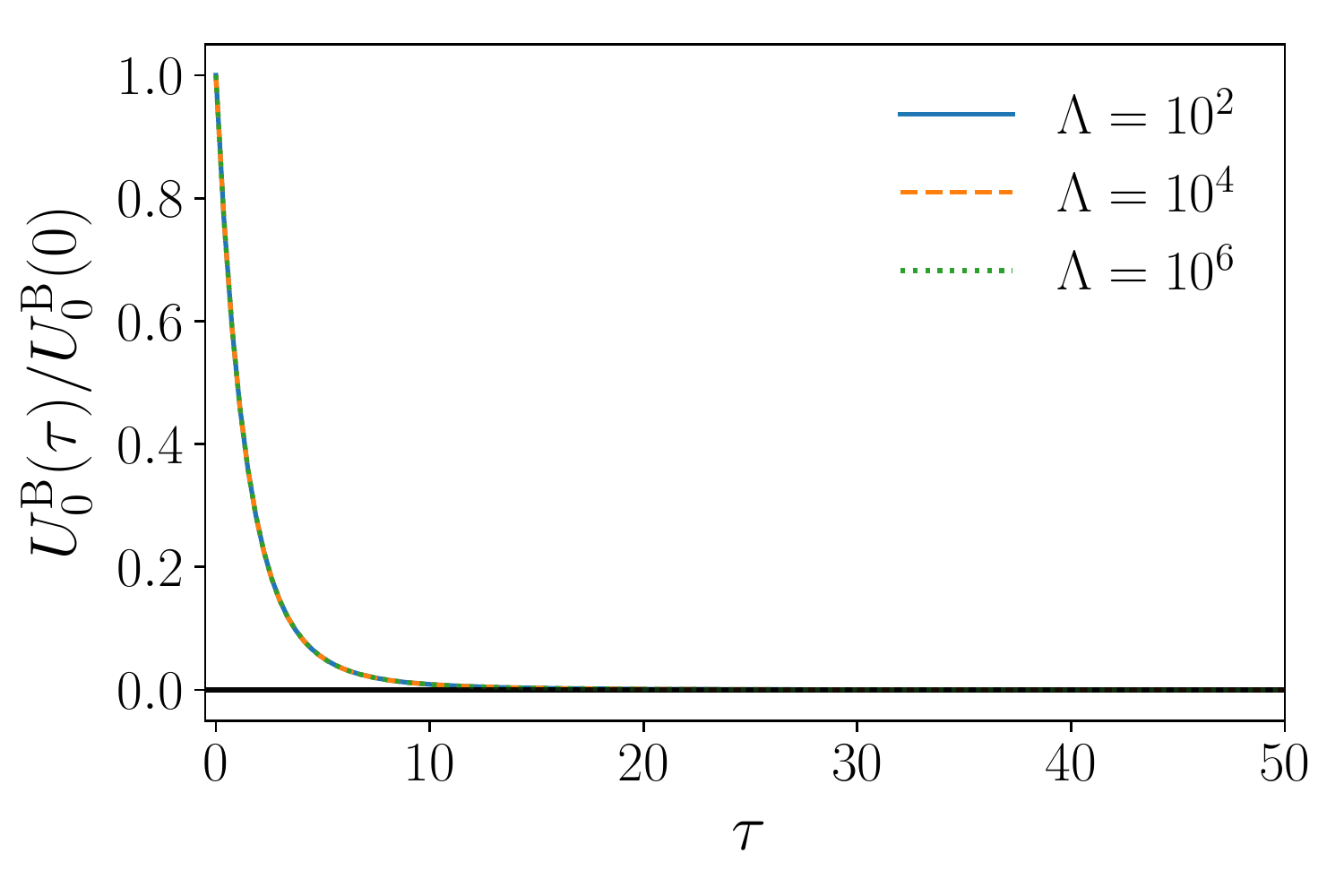}
	\includegraphics[width=0.3\textwidth,clip]{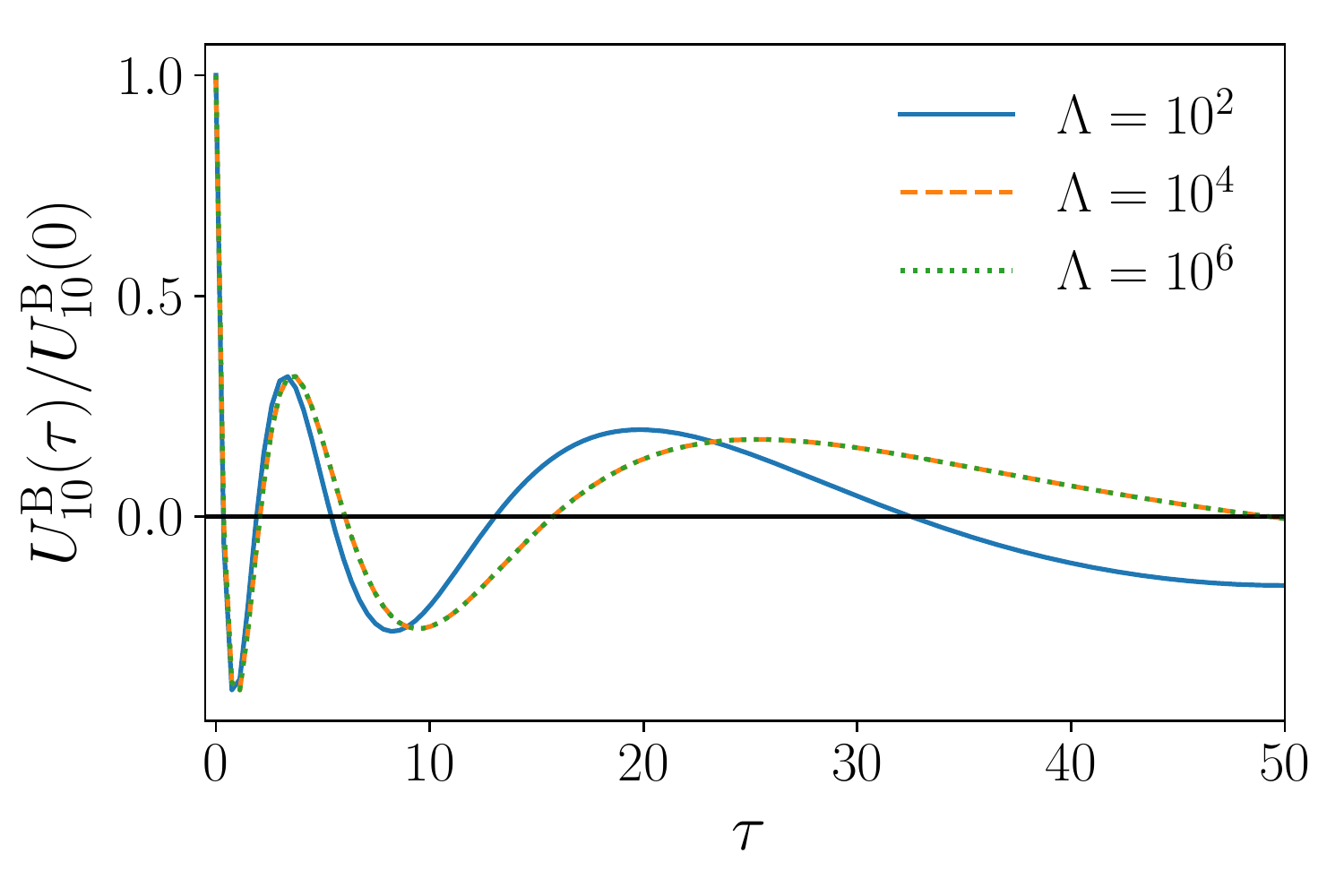}
	\includegraphics[width=0.3\textwidth,clip]{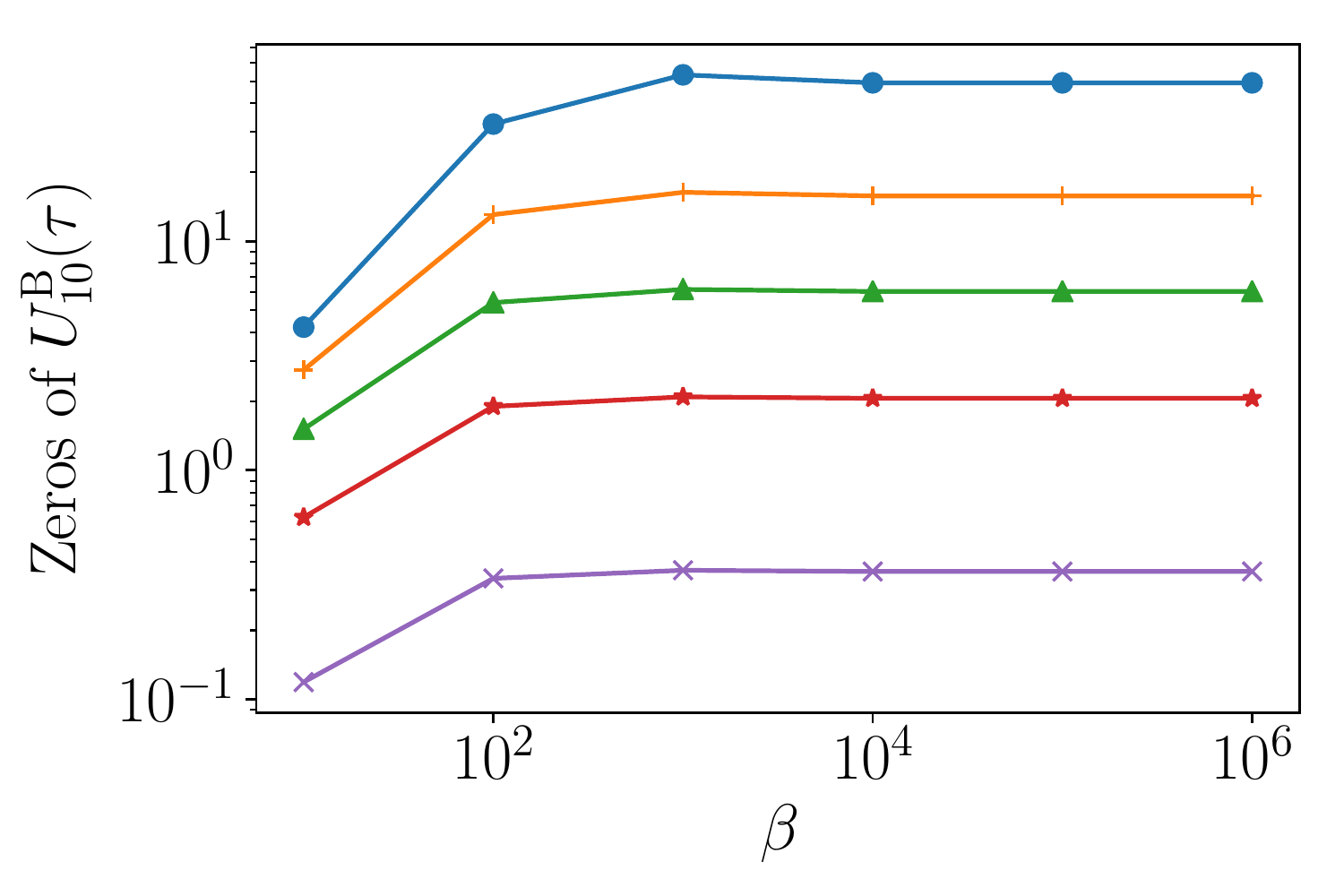}
	\caption{(Color online)
	$\Lambda$ dependence of IR basis functions $U_0^\alpha(\tau)$ near $\tau = 0$ ($\wmax=1$).
	We also show the $\beta$ dependence of the positions of the zeros in the interval $[0,\beta/2]$.
	In the third figure of (a), the broken lines denote $\propto \beta$ and $\propto \beta^{0.1}$, respectively.
	}
	\label{fig:ulx}
\end{figure}

In this section, we study the properties of the IR basis functions to gain an intuitive understand of the outstanding performance of the IR.
For this purpose, we first analyze the relation between the singular values and the dimensionless parameter $\Lambda = \beta\wmax$. 

We plot $\sla/s_0^\alpha$ computed for several values of $\Lambda$ in Fig.~\ref{fig:sl}.
For both of fermions and bosons,
$\sla/s_0^\alpha$ vanishes exponentially at large $l$ as reported in the previous study~\cite{Shinaoka:2017ix}.
The decay becomes slower only slightly even when $\Lambda$ is increased by several orders of magnitude.
An interesting finding for boson is that $\slB/s_0^\mathrm{B}$ seem to converge to a single curve at large $\Lambda$.
See Appendix~\ref{appendix:scaling} for the detailed analysis of the asymptotic behavior at $l=\infty$.

We now define $\Nas$ as the number of singular values above a certain cutoff.
Equation~(\ref{eq:gl}) indicates that $\Nas$ gives an upper bound for $\NaIR$.
Figures~\ref{fig:sl-cut}(a) shows the $\Lambda$-dependence of $\NFs$ computed for the cutoff values of $10^{-5}$ and $10^{-15}$.
It turns out that $\NFs$ increases roughly logarithmically as $\Lambda$ increases.
This indicates that the logarithmic dependence of $\NFIR$ on $\wmax$ and $\beta$ found in Figs.~\ref{fig:Lambda-dep} and \ref{fig:scaling} is a general feature of the fermionic IR basis.
Figure~\ref{fig:sl-cut}(b) shows the results for bosons.
In clear contrast to fermions, $\NBs$ grows more slowly and even becomes saturated at large $\Lambda$.
This trend is seen more clearly for the cutoff value of $10^{-5}$.
This is a consequence of the convergence of $s^\mathrm{B}_l/s^\mathrm{B}_0$ to the single curve as $\Lambda\rightarrow +\infty$.
This result indicates that the saturation of $\NBIR$ is a model-independent behavior of the bosonic IR.

To understand the physical meaning of these observations,
we analyze the IR basis functions $V^\alpha_l(\omega)$.
These basis functions are related to $U^\alpha(\tau)$ through the integral equation~(\ref{eq:integral-UV}).
Figure~\ref{fig:vly}(a) shows the $\beta$ dependence of $V_l^\mathrm{F}(\omega)$ for $\wmax=1$.
As reported in the previous study\cite{Shinaoka:2017ix},
there is a peak at $\omega=0$ in $V_l^\mathrm{F}(\omega)$,
which becomes shaper as $\beta$ is increased.
For $l>0$, there are additional peaks and zeros away from $\omega=0$.
To see how these features scale with $\beta$,
we show the $\beta$ dependence of the positions of the zeros in Fig.~\ref{fig:vly}(a).
The zeros near $\omega=0$ scales roughly as $O(\beta^{-a})$ with $a \simeq 1$.
The exponent are smaller for zeros away from $\omega=0$.
As a result, their positions depend on $\beta$ more weakly than those near $\omega=0$.
We observed the same trend for other values of $l$ (not shown).
For bosons,  as $\beta$ is increased, $V_l^\mathrm{B}(\omega)$ converges in the whole region of $\omega$ for each $l$ [see Fig.~\ref{fig:vly}(b)].
Another clear distinction from the fermionic case is that $V_l^\mathrm{B}(0)$ vanishes as $\Lambda\rightarrow + \infty$.

In Fig.~\ref{fig:ulx}, we show how these differences in $V_l^\alpha(\omega)$
affect the $\beta$ dependence of $U^\alpha_l(\tau)$.
The fermionic basis functions $U_l^\mathrm{F}(\tau)$ show a substantial $\beta$ dependence even up to $\beta=10^6$, and their zeros scale as power laws of $\beta$.
In contrast, $U_l^\mathrm{B}(\tau)$ and their zeros converges quickly to a single curve as $\beta$ is increased.
Let us remind that all the zeros of the Legendre basis functions scale as $O(\beta)$
(see Fig.~\ref{eq:LG-exp}).
This behavior does not match either the fermionic one or the bosonic one.

We now discuss the physical meanings of these properties of the IR basis functions.
For fermions, the $O(T)$ scaling near $\omega=0$ plays a substantial role in describing low-energy phenomena in a fermionic system efficiently.
For instance, in a Fermi-liquid state, relevant physics takes place in the vicinity of the Fermi level whose width in energy scales as $O(T)$.
On the other hand, the weaker scaling away from $\omega=0$ allows the basis functions to retain the capability of describing features in the high-frequency part of the spectrum even at low $T$.
These results indicate that the fermionic IR basis functions involve energy scale hierarchy.

On the other hand, the $\beta$ dependence of the bosonic basis functions can be understood as follows.
Let $G^\mathrm{B}(\tau)$ be the physical correlation function corresponding to a spin or charge susceptibility. 
Then, its integral over $\tau$ gives the corresponding static susceptibility as $\chi = \int_0^\beta d \tau G^\mathrm{B}(\tau)$.
If there is no ground-state degeneracy, $\chi$ converges to a finite value as $\beta \rightarrow +\infty$.\footnote{If the ground state is degenerate, the spin correlation function involves a constant contribution which does not depend on $\tau$. This constant term is not expanded into $U_l^\alpha(\tau)$ compactly. Therefore, one should treat this constant term separately. We refer the interested reader to a related discussion in Section III A of Ref.~\onlinecite{shinaoka-unpublished}.}
This requires that, as $\beta$ is increased, $G^\mathrm{B}(\tau)$ should not change its form around $\tau=0$ and $\beta$ and vanish around $\tau=\beta/2$.
The bosonic IR basis functions clearly satisfy these physical requirements. 

Let us emphasis that 
the construction of the IR basis functions takes into account these physical facts through the definition of $K^\alpha(\tau, \omega)$.
These may explain why the IR provides a more compact representation at low temperature than the conventional orthogonal polynomial representation with no physical ground.

\section{Summary and discussion}
In summary, we studied the performance of the representation of the single-particle Green's function using the IR basis functions.
This was accomplished by developing an algorithm for computing the IR basis functions of arbitrary high degree.
Our findings obtained in Sec.~\ref{sec:model-analysis} are summarized as follows.
First, when the frequency cutoff $\wmax$ is larger than the width of the spectrum $\Omega$,
The number of basis functions required to represent $G^{\alpha}(\tau)$ within a given tolerance, $\NaIR$ ($\alpha=$F, B), grows only logarithmically with $\wmax$.
This indicates that the choice of $\wmax$ only slightly influences convergence, which is beneficial in practical applications.
Second, as $\beta$ is increased with $\wmax$ fixed,
$\NFIR$ grows as $O(\log \beta)$ for fermions, while $\NBIR$ converges to a finite value for bosons.
Thus, $\NaIR$ grows more slowly than any power of $\beta$, indicating potential superiority to the existing technology.

In Sec.~\ref{sec:detail}, we analyzed the properties of the IR basis functions in detail. 
By clarifying the distribution of the singular values $s_l^{\alpha}$,
we showed that the above-mentioned weak dependences of $\NaIR$ on $\wmax$ and $\beta$ are a general property of the IR.
We revealed the existence of energy scale hierarchy in the fermionic basis functions.
We further showed that the IR basis functions for boson correctly capture 
the characteristic features of physical $G^\mathrm{B}(\tau)$, e.g., convergence of the static susceptibility at low temperatures.

The IR basis functions may be useful not only for storing the imaginary-time Green's function but also for performing many-body calculations.
The problem of computation time and memory consumption is particularly severer for calculations based on theories involing two-particle Green's functions such as Bethe-Salpeter equation~\cite{Kunes:2011is,Tagliavini-arXiv}, parquet equations~\cite{Valli2015,Li:2016bna}, and diagrammatic extensions of DMFT~\cite{Rohringer-arXiv, Wentzell-arXiv}.
In Ref.~\onlinecite{Shinaoka:2017ix}, some of the authors and co-workers analyzed the spectral representations of two-particle Green's functions.
They showed that two-particle Green's functions can be expanded compactly in terms of the IR basis functions for the single-particle Greens' function.
In addition, it was shown that the convergence properties of the two-particle Green's functions are essentially identical to those of the single-particle Green's function.
Together with this,
the present results indicate that the data size of the two-particle Green's function increases more slowly than any power of $\beta$.
The IR will enable efficient treatment of these diagrammatic equations at two-particle level.

\begin{acknowledgments}
	We are grateful to Shintaro Hoshino, Masayuki Ohzeki, Manaka Okuyama, Takahiro Misawa and Kazuyoshi Yoshimi for fruitful discussions.
	NC and HS were supported by JSPS KAKENHI Grant No. 16H01064 (J-Physics), 16K17735.
	JO was supported by JSPS KAKENHI Grant No. 26800172, 16H01059 (J-Physics).
\end{acknowledgments}

\appendix

\section{Method for computing basis functions}\label{appendix:method}
In this Appendix, we describe an efficient method for solving the integral equation (\ref{eq:int-solve}).
We note that a solution of the integral equation is either even or odd with respect to $x$ and $y$ since $k^\alpha(x,y) = k^\alpha(-x, -y)$.
We compute even and odd solutions separately by solving the two integral equations
\begin{align}
	s_i u^\mathrm{even}_i(x) &= \int_0^1 dy k^\mathrm{even}(x, y) v_i^\mathrm{even}(y),\label{eq:int-even}\\
	s_i u_i^\mathrm{odd}(x) &= \int_0^1 dy k^\mathrm{odd}(x, y) v^\mathrm{odd}_i(y)\label{eq:int-odd}
\end{align}
under the orthonormal condition $\int_0^1 u_i(x) u_j(x) dx = \int_0^1 v_i(y) v_j(y) dy = \delta_{ij}$.
Here, we defined $k^\mathrm{even} \equiv k^\alpha(x, y) + k^\alpha(x,-y)$ and $k^\mathrm{odd} \equiv k^\alpha(x, y) - k^\alpha(x,-y)$.
Hereafter, we drop the index $\alpha$ for statistics for simplicity.
Once these two equations are solved,
the solutions of the original equation (\ref{eq:int-solve}) are constructed by sorting singular values $s_i^P$ in decreasing order ($P$ = even, odd).

According to the Bobnov-Galerkin method~\cite{al:2007vm},
we expand $u^P_i(x)$ and $v^P_i(y)$ in terms of complete orthonormal coordinate functions $\{\phi_{s,p}(x)\}$, $\{\tilde{\phi}_{s,p}(y)\}$ as
\begin{align}
	&u^P_i(x) = \sum_{s,p}^N u^P_{i,s,p}\phi_{s,p}(x),\label{eq:u-exp}\\
	&v^P_i(y) = \sum_{s,p}^{\tilde{N}} v^P_{i,s,p}\tilde{\phi}_{s,p}(y),\label{eq:v-exp}
\end{align}
where $\int_0^1 dx \phi_{i,s}(x) \phi_{j,s^\prime}(x) = \int_0^1 dy \tilde{\phi}_{i,s}(y) \tilde{\phi}_{j,s^\prime}(y) = \delta_{ij}\delta_{ss^\prime}$.
Our coordinate functions are defined on multi-domains as
\begin{align}
	& \phi_{s,p}(x) \equiv \nonumber \\
	&
	\begin{cases}
		\sqrt{\frac{2}{x_{s+1}-x_s}}
		\tilde{P}_p\left(\frac{x+1}{2}(x_{s+1}-x_s) + x_s\right)~(x_s \le x \le x_{s+1})\\
		0~(\mathrm{otherwise})
	\end{cases},\\
	& \tilde{\phi}_{s,p}(y)\equiv \nonumber \\
	&
	\begin{cases}
		\sqrt{\frac{2}{y_{s+1}-y_s}}
		\tilde{P}_p\left(\frac{y+1}{2}(y_{s+1}-y_s) + y_s\right)~(y_s \le y \le y_{s+1})\\
		0~(\mathrm{otherwise})
	\end{cases}
\end{align}
for $p=0, \cdots, N_p-1$.
In the present study, we use $N_p=10$.
The index $s$ runs from 1 to $N_s$ for $x$, while it runs from 1 to $\tilde{N}_s$ for $y$.
We have $N_pN_s$ coordinate functions for $x$ and $N_p\tilde{N}_s$ ones for $y$, respectively.
The end points of the domains are in ascending order: $x_1 (= 0) < x_2 < \cdots < x_{N_s-1} < x_{N_s} (=1)$ and $y_1 (= 0) < y_2 < \cdots < y_{\tilde{N}_s-1} < y_{\tilde{N}_s} (=1)$.
For convenience, we have defined
\begin{align}
	\tilde{P}_p(x) \equiv \sqrt{\frac{2l+1}{2}} P_p(x)
\end{align}
so that $\int_{-1}^1 dx \tilde{P}_p(x) \tilde{P}_{p^\prime}(x) = \delta_{pp^\prime}$.
We will explain how the distribution of the domains is optimized in an adaptive way below.

We define a combined index of section $s$ and polynomial order $p$ as $k\equiv (s,p)$.
For a given distribution of domains,
we compute the expansion coefficients $u^P_{i,k}$ and $v^P_{i,k^\prime}$ by performing the SVD of the matrix representation of the two kernels
\begin{align}
	\bK_P &=  \bU_P \bS_P (\bV_P)^T,\label{eq:svd}
\end{align}
where
\begin{align}
	(\bK_P)_{kk^\prime} &\equiv \int_0^1 dx \int_0^1 dy\hspace{0.5em}\phi_k(x) k^P(x,y) \tilde{\phi}_{k^\prime}(y).\label{eq:Kij}
\end{align}
$\bS$ is a diagonal matrix with diagonal elements of $s^P_i$.
The expansion coefficients are given by the column vectors of $\bU_P$ and $\bV_P$.
Note that the solutions satisfy the orthonormal condition by construction.

We determine an optimal distribution of domains
by monitoring the truncation error in the following procedure.
First, we compute the positions of the zeros of $u_i^P(x)$ and $v_i^P(y)$ roughly by the SVD of the kernels discretized for $x$ and $y$, respectively.
We use them as breaking points of the interval to generate an initial set of domains.
Then, following the above procedure,
we compute $u_i^P(x)$ and $v_i^P(y)$.
Then, for each domain, we estimate the truncation error of the most rapidly oscillating function, i.e., the solution for the smallest singular value of interest.
If convergence is not reached at some domains, we split them into halves.
Once the distribution of domains is updated,
we compute more accurate solutions $u_i^P(x)$ and $v_i^P(y)$.
We repeat this procedure until the truncation error becomes sufficiently small for all domains.

\section{Convergence properties of the Legendre basis}\label{appendix:legendre}
The power-law scaling can be understood as follows.
We assume that the imaginary-time resolution of $P_l(x(\tau))$ is determined by the closest zero to $\tau=0$, which $1-x_1$ scales as $O(\beta/l^2)$ at large $l$.
On the other hand, the time scale of the decay of $G^\alpha(\tau)$ near $\tau=0$ is $O(1/\wmax)$ ($\wmax=1$ in the present case).
From the condition that these two time scales match, i.e., $\beta/l^2 \propto 1$,
we obtain the scaling relation
\begin{align}
\NaL(\delta) &\propto \sqrt{\beta},
\end{align}
which explains the power-law behavior.

\section{Convergence of spectral function}\label{appendix:spectrum}
\begin{figure}
    \centering
    \includegraphics[width=0.4\textwidth,clip]{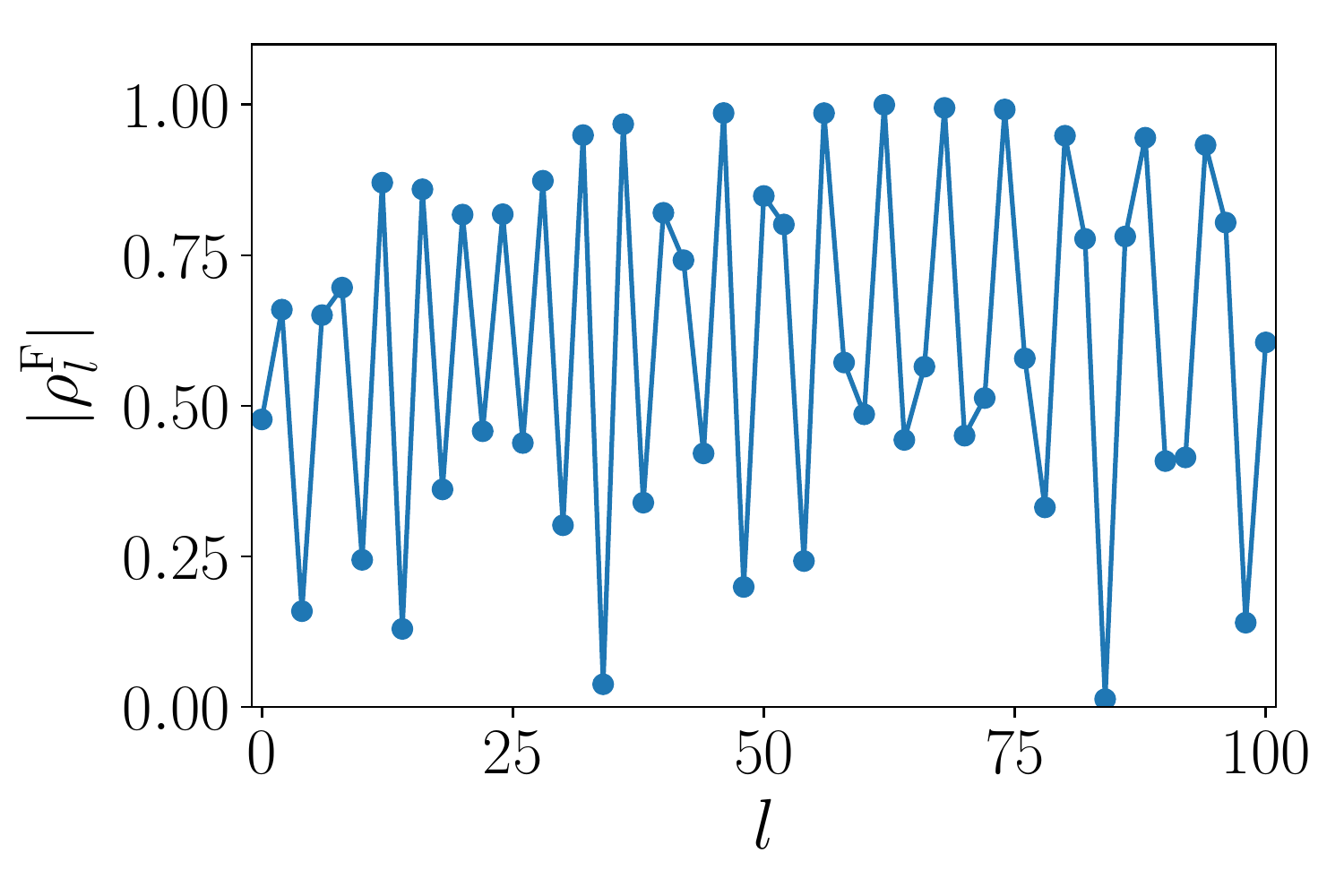}
    \includegraphics[width=0.4\textwidth,clip]{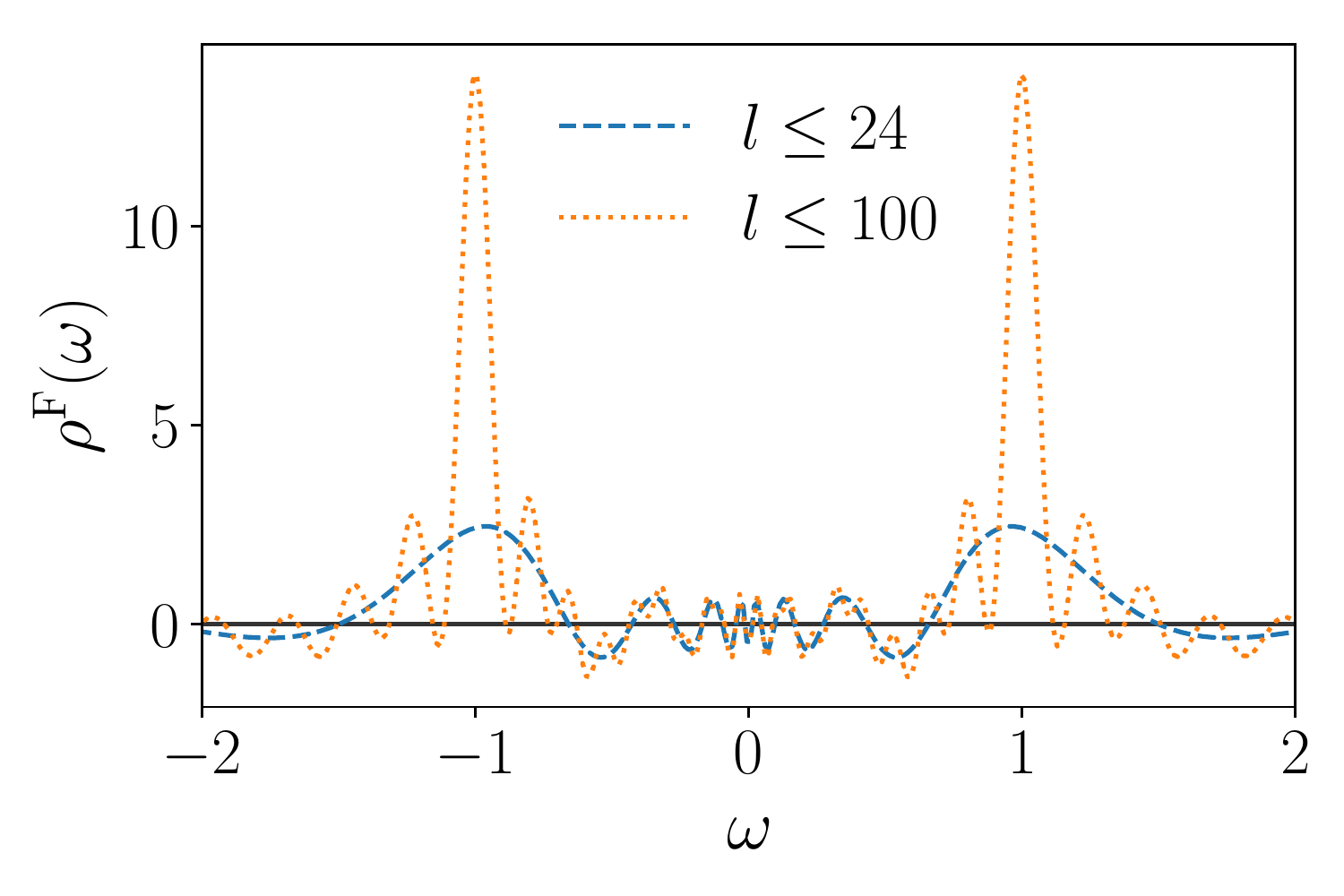}
	\caption{
		(Color online)
		Expansion coefficients of the spectral function $\rho^\mathrm{F}_l$ for the insulating model and reconstructed spectral function from coefficients for $l\le 24$ and $l \le 100$.
		In the upper panel, we show data only for even $l$.
	}
	\label{fig:conv-rho}
\end{figure}

Figure~\ref{fig:conv-rho} shows the expansion coefficients $\rho^\mathrm{F}_l$ in terms of $V^\mathrm{F}_l(\omega)$ for fermion ($\beta=10^2$ and $\wmax=10$).
We also plot the spectral functions $\rho^\mathrm{F}(\omega)$ reconstructed from the coefficients for $l \le 24$ and $l \le 100$, respectively.
Although the imaginary-time Green's function $G^\mathrm{F}(\tau)$ is described very accurately by $U^\mathrm{F}_l(\tau)$ for $l\le 24$ [see Fig.~\ref{fig:scaling}(b)],
the expansion coefficients $\rho^\mathrm{F}_l$ do not even decay for $l \le 100$.
As a result, the $\rho^\mathrm{F}(\omega)$ converges very slowly.
One can find a similar analysis for a more realistic model in Ref.~\onlinecite{Otsuki:2017er} (see Fig.~5).

\section{Asymptotic behavior of IR basis functions}\label{appendix:scaling}
\begin{figure}
	\centering
	\includegraphics[width=0.4\textwidth,clip]{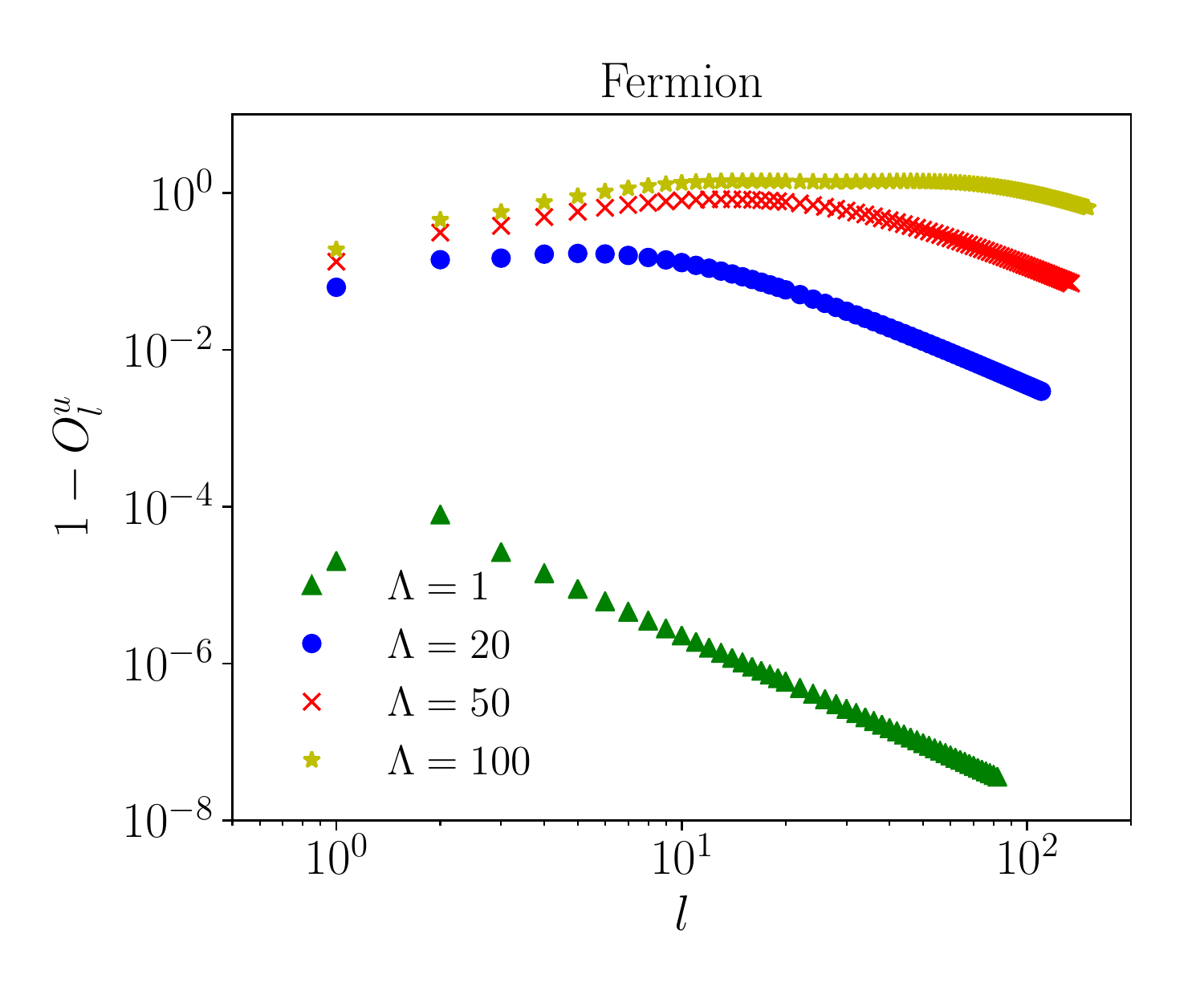}
	\includegraphics[width=0.4\textwidth,clip]{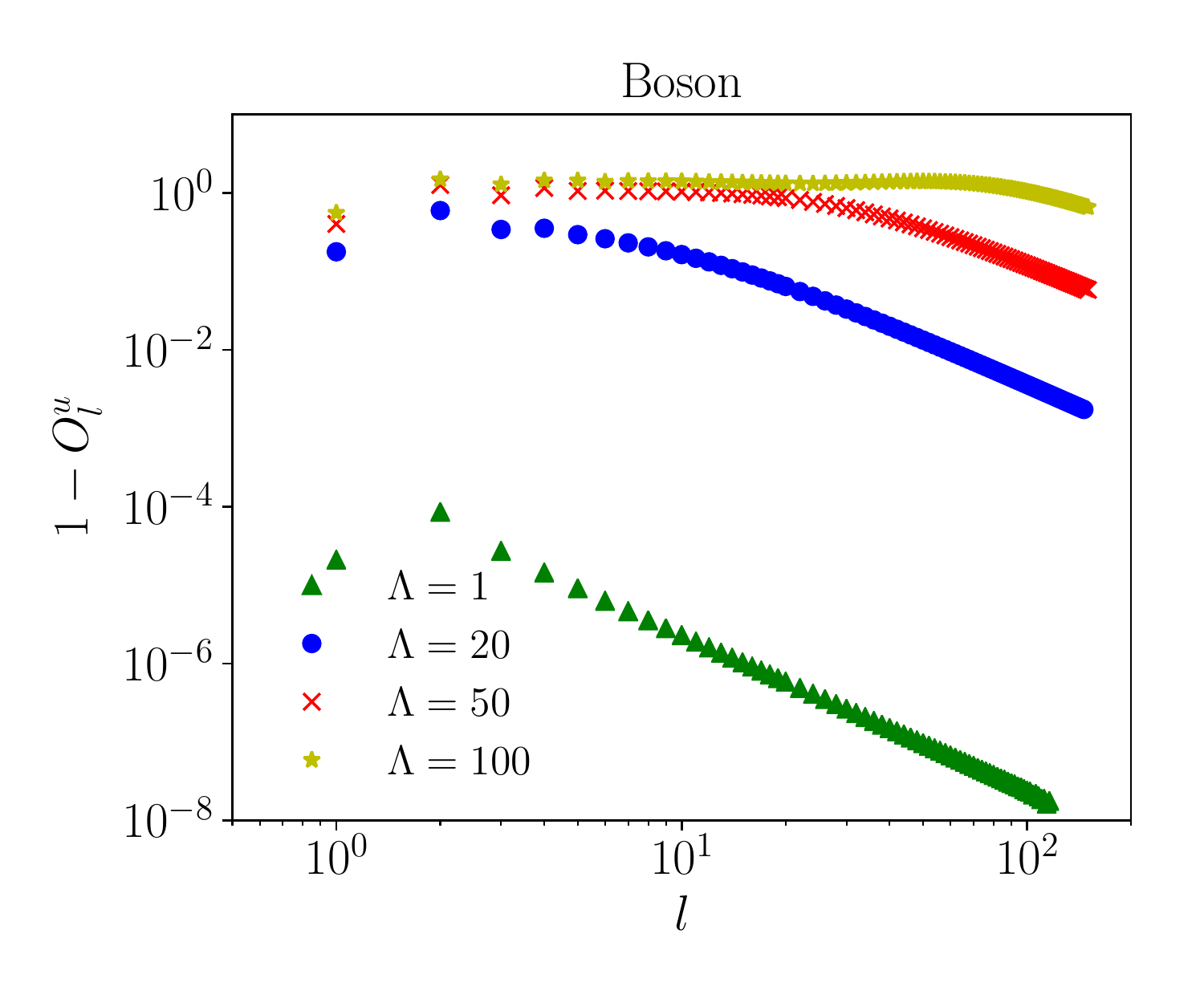}
	\caption{(Color online)
		Difference of the overlap between $u_l^\alpha(x)$ and Legendre polynomials from 1.
	}
	\label{fig:overlap-ulx}
\end{figure}
\begin{figure}
	\centering
	\includegraphics[width=0.4\textwidth,clip]{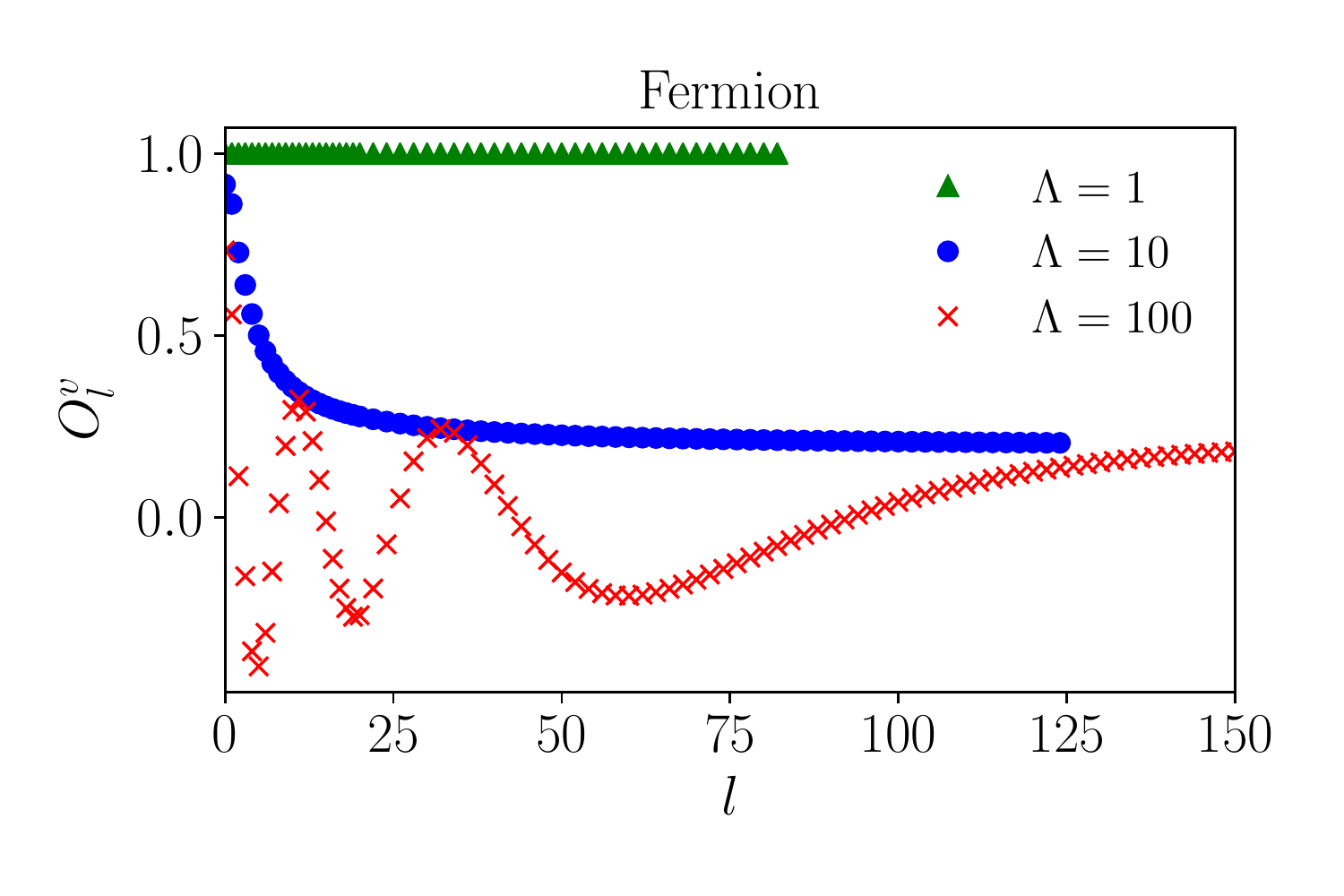}
	\includegraphics[width=0.4\textwidth,clip]{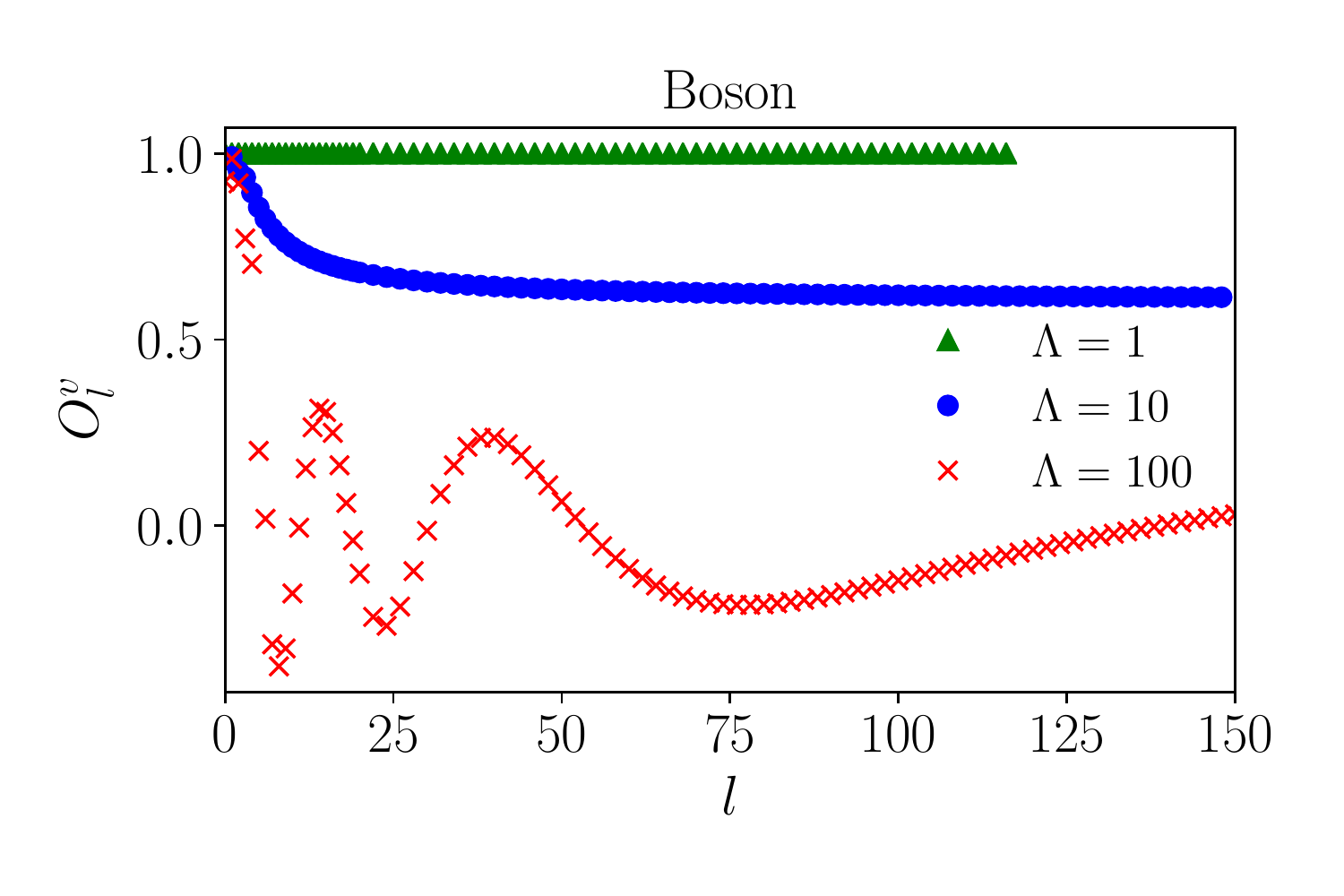}
	\caption{(Color online) Overlap between $v_l^\alpha(y)$ and Legendre polynomials.
	}
	\label{fig:overlap-vly}
\end{figure}
In the previous study~\cite{Shinaoka:2017ix},
it was shown that $u_l^\alpha(x)$ and $v_l^\alpha(y)$ converge to Legendre polynomials up to normalization factors as $\Lambda \rightarrow 0$.
This motivates us to study the overlap between the IR basis functions and Legendre polynomials for finite values of $\Lambda$.
We define the overlap between these two basis functions as
\begin{align}
O^u_{l} &= \sqrt{2l+1}\int_{-1}^1 dx~u_l^\alpha(x) P_l(x),
\end{align}
where $|O^u_l| \le 1$ by definition.
Figure~\ref{fig:overlap-ulx} shows results computed for $\Lambda \le 100$.
An interesting observation is that $1-O^u_l$ vanishes as power laws as $l$ is increased.
This indicates that $u_l^\alpha(x)$ converges to $P_l(x)$ as $l\rightarrow +\infty$ for a fixed value of $\Lambda$.
We can define the crossover point by the onset of the power-law decay.
The crossover point however  grows rapidly as $\Lambda$ is increased.
In particular, for $\Lambda = 100$, the residual starts to decay as a power law only at $l\simeq 100$,
where $s_l^\alpha/s_0^\alpha$ are much less than $10^{-10}$.
Thus, IR basis functions in the smaller-$l$ region of physical interest are substantially different from Legendre polynomials.

We define the overlap between $v_l^\alpha(y)$ and Legendre polynomials as
\begin{align}
O^v_{l} &\equiv \sqrt{2l+1}\int_{-1}^1 dy~v_l^\alpha(y) P_l(y).
\end{align}
Figure~\ref{fig:overlap-vly} shows results computed fro $\Lambda=1$, 10, 100.
In contrast to $u_l^\alpha(x)$, $v_l^\alpha(y)$ do not seem to converge to Legendre polynomials as $l$ is increased.
This asymmetry between $u_l^\alpha(x)$ and $v_l^\alpha(y)$ may originate from the denominator of the kernel in Eqs.~(\ref{eq:kernel-F-xy}) and (\ref{eq:kernel-B-xy}).

\begin{figure}
	\centering
	\includegraphics[width=0.4\textwidth,clip]{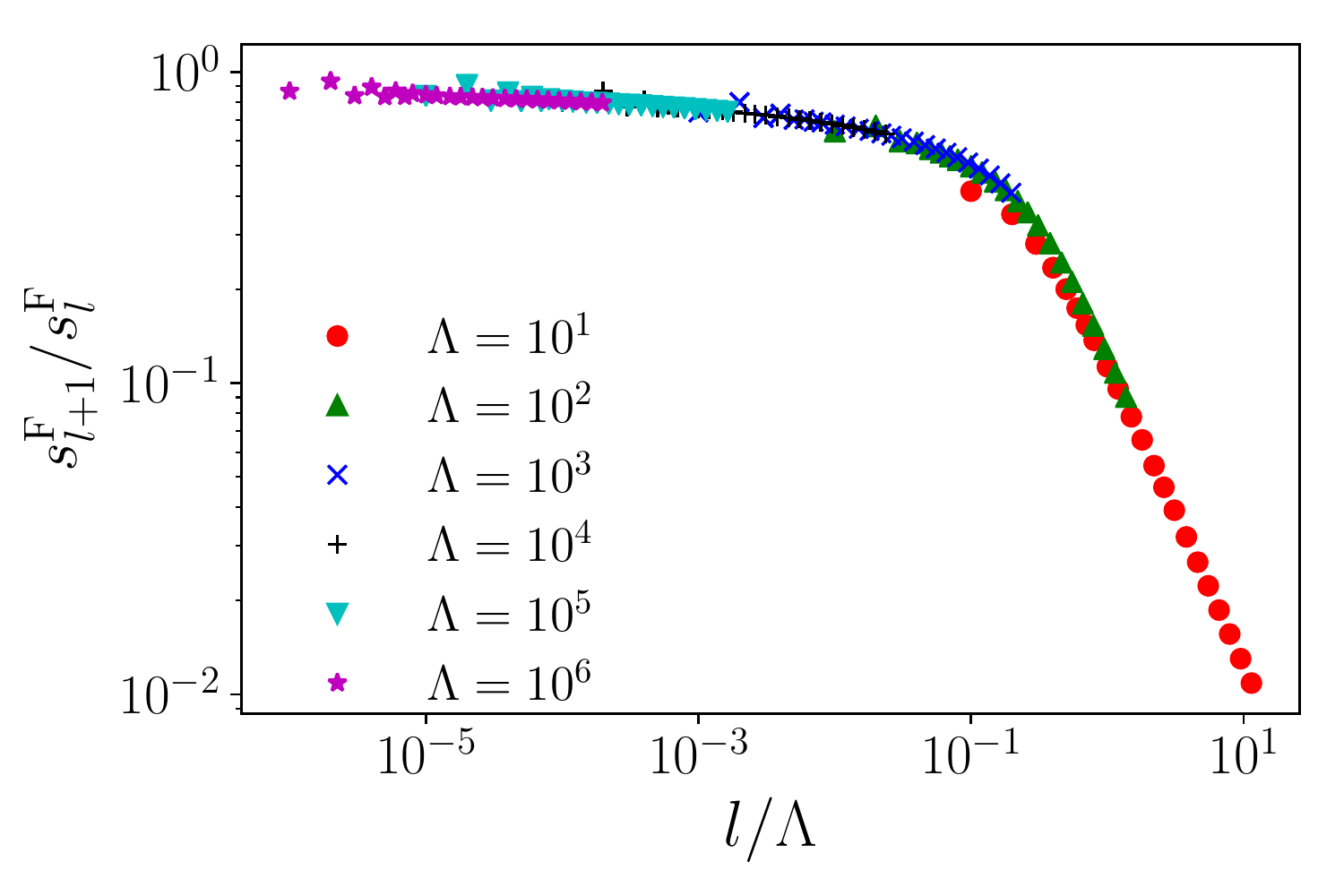}
	\includegraphics[width=0.4\textwidth,clip]{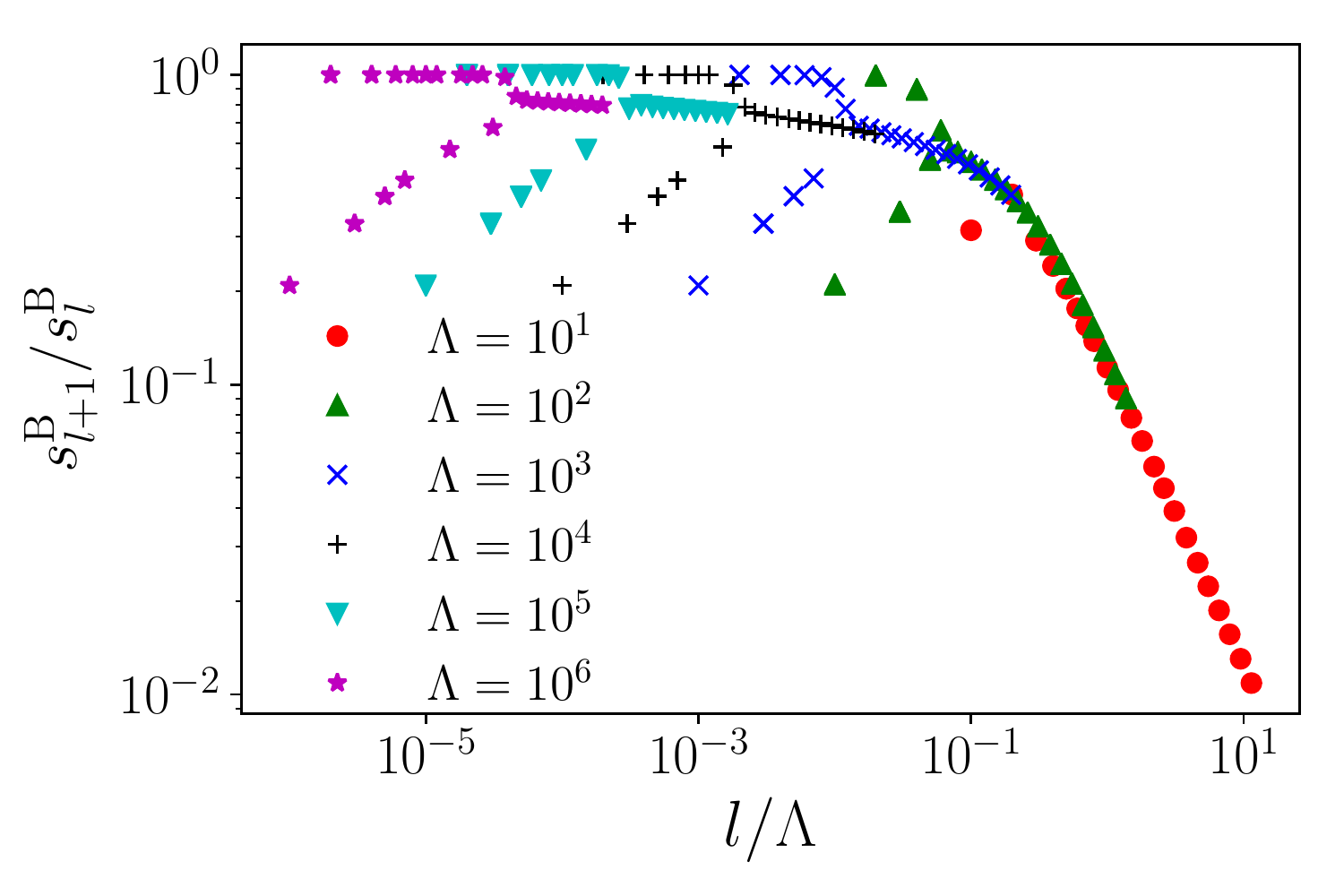}
	\caption{
		(Color online) Scaling plot of decay rate of singular values.
	}
	\label{fig:sl-scaling}
\end{figure}

The crossover to Legendre polynomials may be related to a scaling behavior of singular values.
Figure~\ref{fig:sl-scaling} shows a scaling plot where the decay rate of singular values, $s_{l+1}/s_l$, is plotted against $l/\Lambda$.
Interestingly, the data for all values of $\Lambda$ seem to collapse on a single curve.
For $l/\Lambda \lesssim 1$, the decay rate is almost independent of $l/\Lambda$.
This means that the singular values decay roughly exponentially as $s_l^\alpha \propto e^{-c l}$, where $c$ is a positive constant.

On the other hand, $l/\Lambda \gtrsim 1$, the decay rate scales linearly with $l/\Lambda$ in logarithmic scale.
This means that the decay rate increases as $l$ is increased as $s_{l+1}/s_l \propto l/\Lambda$ in this asymptotic region.
This indicates the scaling relation: $s_l^\alpha \propto \frac{\Lambda^l}{l!}$.

The crossover point grows rapidly as $l_\mathrm{crossover} \simeq 0.1\Lambda$.
Although we were not able to compute singular values for large enough $l$ to reach the crossover point for $\Lambda > 1000$.
the scaling relation seems to hold for a wide parameter region of $\Lambda$.
A remaining issue is how the two asymptotic behaviors of singular values and $u^\alpha_l(x)$ are related.
Although we naively expect that they coincide with each other, a more detailed analysis is left for a future study.

\bibliography{ref_shinaoka_abbrev,ref_aux}

\end{document}